\documentclass[sigconf]{acmart}
\AtBeginDocument{%
  }

\copyrightyear{2026}
\acmYear{2026}
\setcopyright{cc}
\setcctype{by}
\acmConference[CHI '26]{Proceedings of the 2026 CHI Conference on Human Factors in Computing Systems}{April 13--17, 2026}{Barcelona, Spain}
\acmBooktitle{Proceedings of the 2026 CHI Conference on Human Factors in Computing Systems (CHI '26), April 13--17, 2026, Barcelona, Spain}
\acmPrice{}
\acmDOI{10.1145/3772318.3791049}
\acmISBN{979-8-4007-2278-3/2026/04}

\usepackage{multirow}
\usepackage{soul}

\newcommand{\cc}{\texttt{correct-consistent}}
\newcommand{\ci}{\texttt{correct-inconsistent}}
\newcommand{\ic}{\texttt{incorrect-consistent}}
\newcommand{\ii}{\texttt{incorrect-inconsistent}}
\newcommand{\keyresult}[1]{\textbf{#1}}
\newcommand{\figref}[1]{Fig.~\ref{#1}}
\newcommand{\tabref}[1]{Table~\ref{#1}}
\newcommand{\secref}[1]{Sec.~\ref{#1}}

\newcommand{\appendixref}[1]{Appendix Sec. ~\ref{#1}}

\begin{document}

\title[Presenting LLMs as Companions Affects People's Mental Capacity Attributions]{Presenting Large Language Models as Companions Affects What Mental Capacities People Attribute to Them}


\author{Allison Chen}
\orcid{0000-0002-0919-1849}
\affiliation{%
  \institution{Princeton University}
  \city{Princeton}
  \state{New Jersey}
  \country{USA}
}
\email{allisonchen@princeton.edu}

\author{Sunnie S. Y. Kim}
\authornote{Currently at Apple}
\affiliation{%
  \institution{Princeton University}
  \city{Princeton}
  \state{New Jersey}
  \country{USA}
}
\email{sunniesuhyoung@gmail.com}

\author{Angel Franyutti}
\authornote{Work done while an REU student at Princeton}
\affiliation{%
  \institution{Rutgers University}
  \city{New Brunswick}
  \state{New Jersey}
  \country{USA}
}
\email{angelnfranyutticintron@gmail.com}

\author{Amaya Dharmasiri}
\affiliation{%
  \institution{Princeton University}
  \city{Princeton}
  \state{New Jersey}
  \country{USA}
}
\email{amayadharmasiri@princeton.edu}

\author{Kushin Mukherjee}
\affiliation{%
  \institution{Stanford University}
  \city{Stanford}
  \state{California}
  \country{USA}
}
\email{kushinm@stanford.edu}

\author{Olga Russakovsky}
\affiliation{%
  \institution{Princeton University}
  \city{Princeton}
  \state{New Jersey}
  \country{USA}
}
\email{olgarus@princeton.edu}

\author{Judith E. Fan}
\affiliation{%
  \institution{Stanford University}
  \city{Stanford}
  \state{California}
  \country{USA}
}
\email{jefan@stanford.edu}

\renewcommand{\shortauthors}{Chen et al.}

\begin{abstract} 
    How might messages about large language models (LLMs) found in public discourse influence the way people think about and interact with these models?
    To explore this question, we randomly assigned participants ($N = 470$) to watch short informational videos presenting LLMs as either machines, tools, or companions---or to watch no video. 
    We then assessed how strongly they believed LLMs to possess various mental capacities, such as the ability to have intentions or remember things. 
    We found that participants who watched video messages presenting LLMs as companions reported believing that LLMs more fully possessed these capacities than did participants in other groups. 
    In a follow-up study ($N = 604$), we replicated these findings and found nuanced effects on how these videos also impact people's reliance on LLM-generated responses when seeking out factual information.
    Together, these studies suggest that messages about LLMs---beyond technical advances---may shape what people believe about these systems and how they rely on LLM-generated responses. \looseness=-1
\end{abstract}




\maketitle

\section{Introduction}

As artificial intelligence (AI) technology advances, many competing messages about these systems have emerged.
Some messages focus on their applications as tools to enhance productivity \citep{dell2023navigating, weber2024significant, openaiIntroducingDeep}. 
Others emphasize how AI can provide a mechanistic understanding of the nature of intelligence \citep{jumper2021highly, wang2023scientific, trinh2024solving}. 
Still others highlight their potential to provide companionship and emotional support \citep{stade2024large, demszky2023using, koulouri2022chatbots, guingrich2023chatbots}.
How do these different messages about AI shape the beliefs people hold about what these systems fundamentally \textit{are} \citep{newman2019essence, gelman2003essential}---like what underlying capacities they might possess \citep{weisman2017rethinking, thellman2022mental, colombatto2024folk, colombatto2025influence}?
On the one hand, people might already have established beliefs about AI and thus be unaffected by messages that promote different ways of conceptualizing AI systems.
On the other, people might be receptive to messages that encourage them to construe AI systems in certain ways---for example, as social partners rather than productivity-enhancing tools \citep{dennett1989intentional, shanahan2023role}. 
It is important to study the consequences of different messages about AI because what people believe about the systems has the potential to impact how they use and rely on them \citep{colombatto2025influence, hindennach2024mindful, thellman2022mental}. \looseness=-1

A large body of work demonstrates that people hold a wide range of beliefs about AI, regarding topics such as the essential nature of the systems or their potential societal impact. 
For example, recent survey studies suggest that people ascribe varying degrees of consciousness to AI systems \citep{colombatto2024folk, scott2023you} and hold a broad range of conceptions about AI \citep{cheng2025tools, kim2023one}. These works show that people liken the systems to entities such as tools or genies to emphasize different capabilities.
Similarly, many works demonstrate that the public holds both optimistic and pessimistic outlooks on how AI will change society \citep{cave2019hopes, cave2019scary, brauner2023does, kelley2021exciting, denia2025ai}.
Yet it remains unclear on the basis of these observational studies what factors give rise to these wide ranges of beliefs and attitudes. \looseness=-1

To address this gap, some work has employed experimental approaches to investigate how specific design decisions impact how people think about \citep{li2024does, yin2019understanding, zhou2025rel} and interact with \citep{kim2024m, bansal2019beyond, buccinca2021trust, zhou2025rel} AI systems, including robots and chatbots. 
These studies have generally found that when AI systems are designed to exhibit more human-like behaviors (e.g., expressions of politeness) or with a human-like appearance, people often judge them as possessing more human-like capabilities and interact with them as if they were also humans \citep{nass1994computers, zhou2025rel, thellman2022mental, li2024does}.
However, while prior work has primarily explored people's beliefs about hypothetical \citep{cave2018portrayals, watson2019rhetoric, noh2025humanlike} or custom-made AI systems \citep{khadpe2020conceptual, pataranutaporn2023influencing}, the effects on people's beliefs about broadly accessible AI technologies remains to be understood.
Further, while understanding the causal impact of design decisions on user experience is valuable for technology developers, these are not the only factors that impact how people think and feel about AI. 
For example, the messages people are exposed to about AI systems might also impact what they believe and know about these technologies \citep{cave2018portrayals, watson2019rhetoric, khadpe2020conceptual, pataranutaporn2023influencing, noh2025humanlike}.
Thus, there is still a need to understand how messages can shape what people believe about the nature of modern AI systems.

Here we fill this need by studying how different messages about large language models (LLMs), a popular type of AI system, impact people's beliefs about these models and how people interact with them.
We focused on three especially prominent messages about LLMs that were inspired by the current discourse around these systems' emerging capabilities. 
The messages were also motivated by Dennett's influential framework which describes a complex system's behavior at three varying levels of abstraction \citep{dennett1989intentional}. 
The first message is that LLMs are essentially \textit{machines}: engineered systems with complex internal mechanics (and no more).
The second is that LLMs are a kind of \textit{tool} for achieving human-defined goals.
The third is that LLMs are sufficiently complex and capable to be considered genuine social partners or \textit{companions}. 
These different messages were each communicated via a 5-minute video, carefully designed to otherwise provide the same information about LLMs, except for these differences in the overarching message. 
We then conducted two pre-registered experimental studies to measure the impact of this video-based message intervention. 
In the first study ($N = 470$), we measured potential consequences on people's beliefs about the mental capacities that LLMs might possess, such as the ability to have intentions, remember things, or feel tired.
In a follow-up study ($N = 604$), we explored how reliable these effects were nine months later amidst a changing AI landscape and what impact the same intervention might have on how people use LLM responses when performing information retrieval tasks.

These studies revealed two key findings.
First, we found that participants who were exposed to the LLMs-as-companions message judged LLMs to possess more fully developed cognitive and emotional capacities than did participants exposed to the other messages. 
Critically, the LLMs-as-machines and LLMs-as-tools messages also influenced how people thought and felt about LLMs (e.g., trust in the outputs of LLMs and overall feelings towards LLMs), but just not their beliefs about these models' latent mental capacities.
Second, we found that presenting LLMs as machines reduced participants' likelihood to rely on LLM responses that were logically inconsistent, suggesting that these messages might help users remain vigilant when interacting with these systems.
Taken together, these findings provide insights concerning the impact of messages about LLMs, but future work is needed to study how they hold in complex real-world settings.
Such experimental studies will be especially valuable for guiding the development of safe and trustworthy LLM-based applications as these softwares become increasingly prevalent. \looseness=-1

\begin{figure*}[t]
    \centering
    \includegraphics[width=\linewidth,trim={0 0 0 0},clip,
    alt={schematic of the experimental design where participants are assigned to one of four video conditions: watch a video presenting large language models (LLMs) as machines, tools, or companions, or watch no video. Each of these conditions are represented graphically in one row with corresponding colors. Below the condition graphics is a simplified version of the survey where participants reported how capable they believe LLMs are of 40 different mental capacities.}]{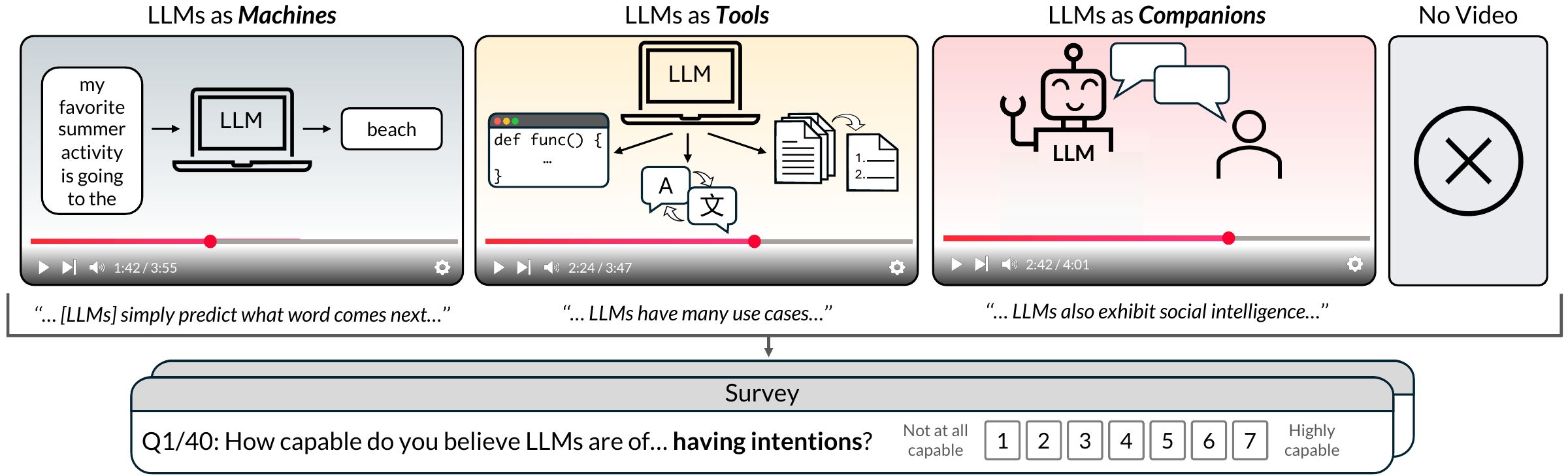}
        \vspace{-0.25in}
        \caption{Overview of Study 1. \textbf{Top}: Participants were randomly assigned to watch one of three short informational videos or to watch no video. \textbf{Bottom:} The survey recorded participants' attributions of 40 mental capacities to LLMs and five additional beliefs about LLMs (not shown): human-likeness, trust in the outputs of, confidence in using, confidence in programming, and overall feelings towards LLMs.}
        \vspace{-0.15in}
\label{fig:schematic}
\end{figure*}

\section{Related Work} \label{related-work}

\subsection{Characterizing Messages about AI}

It is important for the general public to understand the implications of advances in AI systems.
However, it can be challenging for people to achieve this understanding when faced with competing narratives, ranging from more optimistic to more pessimistic ones \citep{cave2019hopes, sartori2023minding, cave2019scary, appio2025ai, denia2025ai}.
Thus simply providing people with \textit{more} information about AI might be insufficient for them to reconcile these different perspectives and develop a clearer understanding of the technology \citep{druckman2025agenda}. 
Instead, it is possible that alternative communication strategies can help provide people with accessible means to understand AI systems.

One such strategy could be to employ metaphors that describe AI systems in human-like terms or as other familiar entities \citep{mcgellin2021stop, carbonell2016role}, such as machines or tools.
However, it is not well understood how these different messages may shape people's beliefs about the nature of the technology.
Currently, both technical experts and laypeople alike describe AI systems using human-like, or anthropomorphic, terms such as ``intelligent'' or ``capable of learning'' \citep{cheng2025tools, carbonell2016role, watson2019rhetoric}.
On the one hand, these anthropomorphic descriptions can make AI systems seem more familiar and approachable, especially because people already tend to ascribe traits like intelligence to AI systems \citep{kim2023anthropomorphic, lee2020perceiving, wang2021towards, gupta2024assistant}.
Yet on the other, these messages can inadvertently lead people to both form unrealistic expectations about an AI system's behavior \citep{cave2018portrayals, salles2020anthropomorphism, mitchell2024metaphors, noh2025humanlike} or develop an overly strong dependence on it \citep{cheng2024one, watson2019rhetoric, thellman2022mental, maples2024loneliness}.

To circumvent these risks, other work has focused on developing educational materials that present AI systems as non-living entities \citep{roe2025reflecting}, like as tools to describe how to use them \citep{mansur2025ai} or simplified machines to explain how they work \citep{carney2020teachable}. 
Then to better understand the space of anthropomorphic and non-anthropomorphic messages about AI, recent works have identified recurring dimensions and patterns across common messages, such as the degree of usefulness or human-likeness each message implies \citep{gupta2024assistant, cheng2025tools}. However, they did not investigate the \textit{causal} impact these common messages might have on what people think about and how people use \textit{current} AI systems, such as LLMs.
Our work goes one step further to investigate how different messages about LLMs might affect people's beliefs about the nature of these systems and how people use them.

\subsection{Attributing Mental Capacities to AI} \label{related-mcs}

Understanding people's beliefs about the nature of AI---such as whether AI systems might possess various mental capacities---can help researchers know \textit{why} people interact with the systems in particular ways \citep{thellman2022mental}.
However, it can be generally challenging for people to accurately report their beliefs about the nature of AI systems. 
To circumvent this problem, human-AI interaction researchers often ask people to report their beliefs about the mental capacities that AI systems might possess, such as having intentions, remembering things, or feeling tired \citep{thellman2022mental, colombatto2024folk, colombatto2025influence, cucciniello2023mind, rossignoli2022attribution}.
This approach is inspired by work in psychology that measured people's judgments about an entity's capacities to identify the important dimensions that constitute its essential nature \citep{weisman2017rethinking, gray2007dimensions, colombatto2024folk, eddy1993attribution, urquiza2015mind, miraglia2023development}---the deep, inherent properties that make it what it is and not something else 
\citep{newman2019essence, gelman2003essential}.
These latent capacities have been argued to be well-captured by few meaningful dimensions
\citep{gray2007dimensions, weisman2017rethinking, malle2019many, tamir2018modeling}, such as \textit{agency} and \textit{experience} \citep{gray2007dimensions} or \textit{body, heart,} and \textit{mind} \citep{weisman2017rethinking}. 
While these works may disagree on which axes are most important, they all derive the dimensions by eliciting people's judgments about a wide range of potential mental capacities, then performing dimensionality reduction on the responses \citep{gray2007dimensions, weisman2017rethinking, colombatto2024folk, malle2019many}. 
In this work, we employ a similar approach to better understand people's beliefs about the essential nature of \textit{LLMs}---a broadly accessible AI system---and, more importantly, to further identify how \textit{messages} about LLMs may shape these beliefs. \looseness=-1

While prior work has demonstrated that people are inclined to ascribe some degree of human-likeness to early AI technology, it is less understood how this extends to today's real-world systems that are more complex and sophisticated, like LLMs. 
A long history of human-computer interaction research suggests that people tend to believe that technology possesses some degree of human-like minds and traits \citep{nass1994computers, kim2012anthropomorphism, weisman2017rethinking, colombatto2024folk}. 
This effect is even greater for systems---such as computers \citep{nass1994computers, kim2012anthropomorphism}, robots \citep{thellman2022mental, manzi2020robot, rossignoli2022attribution, cucciniello2023mind, desideri2021mind}, and chatbots \citep{wischnewski2025attributing, lee2020perceiving, lee2024relationship, khadpe2020conceptual, pataranutaporn2023influencing}---that take on a human-like appearance \citep{thellman2022mental, manzi2020robot}, exhibit polite or honest behavior \citep{thellman2022mental, rossignoli2022attribution, cucciniello2023mind, lee2024relationship}, or are simply described like a person \citep{khadpe2020conceptual, noh2025humanlike}.
Yet, it remains unclear to what extent these findings apply to LLMs, which differ from previous generations of AI systems.
LLMs exhibit more advanced capabilities than earlier systems \citep{bubeck2023sparks, chang2024survey, besta2024graph, wei2022emergent} and are becoming increasingly familiar to the wider public \citep{zhao2023survey, makridakis2023large}. 
However, prior work primarily studied systems that were largely disconnected from people's lives, either hypothetical \citep{shank2018attributions, srinivasan2021algorithms, scott2023you} or created for a specific study \citep{khadpe2020conceptual, pataranutaporn2023influencing, wang2021towards, lee2020perceiving}.
Thus, people's beliefs about the nature of real-world AI systems like LLMs are still relatively understudied. 
The few works that do explore this tend to be observational \citep{colombatto2024folk, colombatto2025influence},
but it is becoming increasingly important to understand the \textit{causal} factors that shape people's beliefs, especially as LLMs become increasingly prevalent in society and the media \citep{karanouh2023mapping, heaton2024chatgpt, makridakis2023large}.
We begin to fill this gap by exploring how different types of messages based on those in the media may shape people's beliefs about what LLMs are, namely how the systems might possess a wide range of mental capacities. \looseness=-1

\subsection{Reliance on AI Systems}

People's beliefs about AI are often intertwined with their reliance behaviors.  
Early work studying reliance tended to focus on AI-assisted decision-making scenarios \citep{klingbeil2024trust}.
They demonstrated that providing information about a system's accuracy \citep{he2023stated, yin2019understanding, yu2019trust, bansal2019beyond} or giving explanations for its prediction \citep{gonzalez2021explanations, vasconcelos2023explanations, kahr2024understanding, bansal2021does, bussone2015role} could increase people's reliance on the systems.
These effects could then be further amplified or moderated by the context of the interaction, such as task demands and individual traits \citep{kuper2025psychological, chen2023understanding, biswas2024impact, swaroop2024accuracy}.

Now as LLMs become more popular, more works are shifting to study reliance on LLMs in a wide range of domains, such as education, medicine, and coding \citep{ vasconcelos2025generation, choudhury2024large, stojanov2024university}. These works demonstrate similar trends: that providing explanations can increase reliance \citep{kim2025fostering, si2024large}, conveying uncertainty can reduce reliance \citep{bo2025rely, kim2024m, zhou2024relying, vasconcelos2025generation}, and individual differences and context can have varying effects \citep{colombatto2025influence, zhou2025rel, zhai2024effects}.
One specific use case that has gained recent attention is understanding how people rely on LLMs to retrieve information \citep{zhai2024large, zhang2025survey, stadler2024cognitive}. 
Since LLMs can be incorrect or inconsistent in both obvious and subtle ways \citep{kim2025fostering, liu2023evaluating, zhai2024large}, it is important to study how people rely on their outputs \citep{kim2025fostering, kim2024m}.
Prior research, specifically \citet{kim2025fostering}, demonstrated that in these settings, people are largely sensitive to the consistency of LLM responses but less so to the correctness.
However, it is not yet known how system-agnostic factors, like \textit{messages} about LLMs, can moderate these effects.
The second part of our work aims to fill this gap
by building upon \citet{kim2025fostering}. 
We study how messages about LLMs might moderate some of these factors, namely the correctness and consistency of LLM explanations, to further shape people's reliance behaviors, and we find nuanced effects.
 
\section{Study 1: Characterizing the Impact of Messages on Beliefs about LLM Mental Capacities}

The main research question motivating our work is: \textit{How do different ways of presenting LLMs---as machines, tools, or companions---affect people's beliefs about the mental capacities that LLMs might possess?}

\subsection{Methods} \label{main-methods}

To explore this question, we conducted a between-subjects experiment (N=470) where participants were randomly assigned to one of four conditions: to watch a video presenting LLMs as machines, tools, or companions, or to watch no video (\figref{fig:schematic}). 
Then all participants took a survey measuring primarily their attributions of mental capacities to LLMs and secondarily their additional beliefs towards LLMs, such as trust in LLM outputs or overall feelings towards LLMs.
This experiment was approved by our institution's Institutional Review Board (IRB) and pre-registered with AsPredicted (\url{https://aspredicted.org/vgdm-gjrm.pdf}).
In this section, we describe our experimental design and procedure, approach to participant recruitment, and statistical analysis methods.

\subsubsection{Developing Different Video Messages About LLMs}
We designed three informational videos presenting LLMs as machines, tools, or companions that each embodied a different and common message about LLMs. 
Each video's content was informed by both current messaging about LLMs and Dennett's hierarchy \citep{dennett1989intentional, thompson2009daniel}.
Dennett's hierarchy articulates theoretical distinctions between different types of explanations  \citep{lombrozo2006functional, lombrozo2012explanation, kelemen2019magic}; it suggests that any complex behavior can be construed at different levels of abstraction by adopting the physical, design, or intentional stance. 
The physical stance focuses on the causal mechanisms of certain behavior \citep{lombrozo2012explanation}, the design stance focuses on what a behavior is intended for \citep{lombrozo2006functional, lombrozo2012explanation}, and the intentional stance focuses on describing said behavior through the lens of agency and intention.
The \textit{machines} video message adopted the physical stance, highlighting the causal mechanisms that enable LLM text generation. 
The \textit{tools} video was based on the design stance, emphasizing what tasks LLMs are suitable for.
Lastly, the \textit{companion} video was founded on the intentional stance, describing LLMs as potential social companions possessing agency and intentions.
Such a theoretical framing is helpful to formally characterize the ways people construe and present LLMs and has practical consequences for understanding how common messages about these systems may shape what people believe about the essential nature of LLMs. \looseness=-1

While real-world messages differ in many aspects, such as context, length, source, and style, we intentionally designed the videos to be as systematically consistent as possible. 
Although the videos may have foregone the complex nuances of real messages and ecological validity, we made this decision to reduce potential confounding factors in our experiment.
Each video contained one distinctive content section, but otherwise all videos were less than 5 minutes long, shared the same introduction, conclusion, and primer about how LLMs learned from data. 
The videos further shared similar graphics, animations, and the same narrator.
In the survey, we presented each video in three parts for maximal engagement; participants were required to stay on each screen for the duration of each part and manually click to move onto the next section.
For further details, links to the videos, and the full scripts, see \appendixref{app-vid}. \looseness=-1

\subsubsection{Measuring Beliefs About LLMs}
To study whether messages about LLMs could influence people's beliefs about the essential nature of these systems, all participants, regardless of whether they watched a video or not, reported how much they attributed 40 mental capacities to LLMs, such as the ability to have intentions, remember things, or feel tired (see \figref{fig:mcs_attitudes} for the full list).
We adopted this methodology from prior psychological work where they observed people's judgments on a wide variety of mental capacities to better understand how people represent an entity's mental nature \citep{weisman2017rethinking, gray2007dimensions, malle2019many, colombatto2024folk}.
Following this line of work, we measured mental capacity attribution using 7-point Likert scales \citep{thellman2022mental, eddy1993attribution, miraglia2023development, weisman2017rethinking}, and our final set of capacities were compiled from two relevant prior works \citep{weisman2017rethinking, colombatto2024folk}. Using a small-scale qualitative pilot study, we identified 40 capacities that spanned a wide variety of relevant characteristics, while minimizing redundancy. Further details can be found in \appendixref{pilot}. 
To additionally explore the extent to which these video interventions impacted other aspects of how people think and feel about LLMs, we also asked participants to answer five secondary questions. Each probed at an additional belief about LLMs using a single 7-point Likert scale: human-likeness of, trust in, confidence in using, confidence in programming, and overall feelings towards LLMs.\looseness=-1

\subsubsection{Data Collection}\label{main-data-collection}

We recruited participants in November 2024 using Prolific, a popular online research platform. 
We targeted U.S.-based adults without specialized knowledge or experience with AI or computer science with a standard sample.
We used both Prolific's pre-screening filters for education, employment, and programming experience as well as custom questions regarding computer science experience (``Did you obtain a degree, are you pursuing a degree, or do you work in an area related to computer science?'') and self-reported knowledge of AI technology (``How would you rate your current knowledge of artificial intelligence (AI)?'').
Participants provided informed consent in accordance with our IRB's requirements, and were compensated \$15 per hour for their participation. \looseness=-1

Our target minimum sample size was $90$ per condition ($360$ total), based on a power analysis conducted using G*Power \cite{erdfelder1996gpower}, with $alpha=0.05$, $power=0.9$, $Cohen's\ d=0.5$ for a two-tailed Mann-Whitney U-test. In total, we recruited 489 participants, then, following our pre-registered exclusion criteria, we excluded 19 participants who spent less than 80 seconds on the mental capacity attribution survey, spent a median time of less than 1 second on each item of the survey, or failed our attention check. The goal of the attention check was to filter out participants who were clearly not paying attention, so we asked participants to select the two options out of the following four that they saw in the mental capacity attribution survey: ``understanding how others are feeling," ``doing computations," ``solving a Rubik's cube," or ``riding a bike."\footnote{The first two are the correct answers.}
After applying these exclusions, we retained data from a total of 470 participants for further analysis ($3.9\%$ exclusion rate): 118 watched no video, 116 watched the video presenting LLMs as machines, 116 as tools, and 117 as companions.
Those who were assigned to watch a video spent a median time of $14$ minutes for the entire survey while the no-video participants spent a median time of $6.5$ minutes.
We encouraged participants to engage with the videos by requiring them to stay on each page for at least the duration of the video and report one thing they learned from the video. However, we ultimately had no control over participants' engagement, and results should be interpreted carefully in light of this limitation.
Further participant demographics can be found in \appendixref{main-demographics}. \looseness=-1

\subsubsection{Analysis Procedure}\label{main-analysis-procedure}

The primary goal of our statistical analysis was to determine how the videos presenting LLMs differentially influenced how participants attributed various mental capacities to LLMs. 
We first categorized the mental capacities using exploratory factor analysis (EFA) as a bottom-up approach to structure to our data. We conducted EFA using the \texttt{fa} package in \texttt{R} and determined to use three factors by minimizing the Bayesian Information Criterion with 5-fold cross validation.
Then, according to our pre-registered analysis plan, we fit the following linear mixed-effects regression model on the data: \texttt{rating $\sim$ video\_condition * category + (1 | p\_id)}. 
In this model, \texttt{rating} was a participant's predicted rating for a specific mental capacity; \texttt{video\_condition} was a categorical variable with four levels, corresponding to our four video conditions with \texttt{no\_video} as the reference level; \texttt{category} represented which factor analysis category this mental capacity belonged to; and \texttt{p\_id} was each participant's unique ID. 
We also employed a similar analysis strategy to assess the impact of these videos on the additional survey questions probing how people thought and felt about LLMs. All analyses achieved a minimum power of $80\%$ when using the \texttt{simr} package in \texttt{R} with $N=1000$ simulations. \looseness=-1

To test for the reliability of the effect of any predictor variable, we report the results of quantitative model comparisons between linear mixed-effects regression models with and without that variable, summarized using standard test statistics: $\chi^2$ or $F$, and $p$. 
We use these regression models also to report quantitative estimates of key outcome variables, namely estimated marginal means, accompanied by 95\% confidence intervals (i.e., $M = Mean \ [CI\ low, CI\ high]$). For implementation details of our analyses, see \appendixref{main-analysis}. \looseness=-1

\subsection{Results} \label{mca-results}

In this section, we report the effect of experimentally manipulating which video participants watched on their beliefs about how LLMs might possess various mental capacities. 

\begin{figure*}[t!]
    \centering
    \includegraphics[width=1.0\linewidth,trim={0 0 0 0},clip,
    alt={Point plots demonstrating results of participants attributions of 40 mental capacities to LLMs. The y-axis represents the ratings (from 1 to 7) and the items are along the x-axis, partitioned based on their categories from factor analysis. Different video conditions are represented with different colors. For the most part, participants in the companion video attributed higher mental capacities to LLMs than other participants.}]{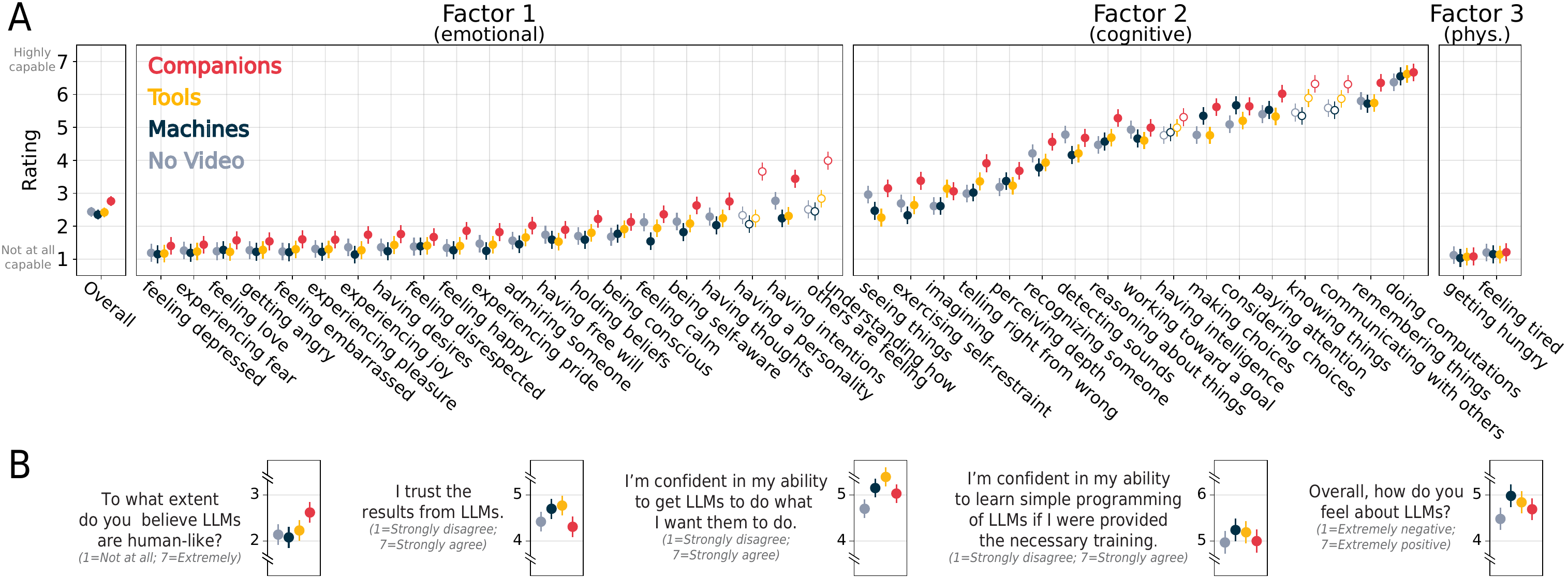}
    \vspace{-0.2in}
    \caption{Study 1 results showing estimated marginal means and 95\% confidence intervals from statistical analyses. 
    All items were measured using 7-point Likert scales. 
    \textbf{A}: Participants' attributions of mental capacities to LLMs. Leftmost facet shows  attributions averaged across all 40 capacities. Right three facets show attributions partitioned by the three exploratory factor analysis categories, which we called \textit{emotional}, \textit{cognitive}, and \textit{physiological}. Open circles indicate capacities that were referenced in the \textit{companion} video. 
    \textbf{B}: Participants' responses of human-likeness of, trust in the outputs of, confidence in using, confidence in programming, and feelings towards LLMs.\looseness=-1} 
    \vspace{-0.15in}
\label{fig:mcs_attitudes}
\end{figure*}

\subsubsection{Overall Impact of Messages on Attributions of Mental Capacities}

To start, we analyzed the overall differences between the no-video baseline and three video conditions in aggregate.
We did not find evidence that participants who watched a video were more likely to attribute mental capacities to LLMs than those who did not watch any video 
((n.s.): $\chi^2 = 2.171$, $p = 0.141$; 
$M_{no-video} = 2.44$ $[2.28, 2.59]$ vs. 
$M_{video} = 2.51$ $[2.42, 2.60]$).
This suggests that regardless of whether they watched a video or not, participants thought LLMs possessed similarly developed mental capacities.  
%
%
%
%
However, we did find that participants who watched a video reported higher \textit{confidence} in their attributions 
($F_{1, 468} = 11.829$, $p < 0.001$; 
$M_{no-video} = 4.47$ $[4.24, 4.71]$ vs. 
$M_{video} = 4.95$ $[4.81, 5.08]$), suggesting that being provided \textit{any} information, regardless of the content, can lead people to be more confident in their beliefs about the mental capacities LLMs might possess. \looseness=-1 

    %
Critically, we found that \textit{which} video participants watched had an effect on their judgments ($\chi^2 = 37.920$, $p < 0.001$).
\keyresult{Specifically, participants who watched the \textit{companion} video attributed more mental capacities to LLMs than the other three groups} (\figref{fig:mcs_attitudes}A; 
$M_{companion} = 2.76$ $[2.61, 2.91]$ vs. 
$M_{other} = 2.40$ $[2.32, 2.49]$). This suggests that these participants believed LLMs to possess more fully developed mental capacities relative to participants in the other conditions. 
Further, we conducted a robustness analysis to omit the five (of 40 total) mental capacities identified by the research team to have been referenced in the video presenting LLMs as companions. Even without these capacities, we observed that this effect persisted ($\chi^2 = 29.987$, $p < 0.001$), suggesting that the participants exposed to the \textit{companion} video message drew inferences about LLM mental capacities that went beyond those mentioned in the video. \looseness=-1

\subsubsection{Differential Impact of Messages on Attributions of Cognitive, Emotional, and Physiological Capacities}

Thusfar, our results demonstrate that the video presenting LLMs as companions can lead people to believe LLMs might possess more fully developed mental capacities.
Next, we performed a finer-grained analysis to understand how people may attribute different categories of capacities to LLMs. 
We used the three emergent categories from our factor analysis, which we informally refer to as ``emotional,'' ``cognitive,'' and ``physiological'' to group our 40 mental capacities (see \figref{fig:mcs_attitudes}A for categorizations).
%
%
%
We first observed a significant main effect of the mental capacity category on participants' attributions ($\chi^2 = 11474.766$, $p < 0.001$). 
In other words, regardless of video condition, all participants tended to report LLMs to be more likely of possessing ``cognitive'' capacities than ``emotional'' or ``physiological'' ones
(``cognitive'': $M = 4.60$ $[4.53, 4.68]$;
``emotional'': $M = 1.75$ $[1.68, 1.83]$;
``physiological'': $M = 1.12$ $[1.01, 1.24]$), consistent with prior work \citep{weisman2017rethinking, colombatto2024folk}. \looseness=-1

We then analyzed the degree to which the video intervention differentially influenced attributions of different categories of capacities. 
Here we observed an interaction ($\chi^2 = 20.774$, $p = 0.002$), where the \keyresult{participants who watched the \textit{companion} video attributed LLMs with more ``emotional'' and ``cognitive'' capacities compared to participants in other groups.
However, all participants attributed similar levels of ``physiological'' capacities} (\figref{fig:mcs_attitudes}A; 
``cognitive'': $M_{companion} = 5.00$ $[4.85, 5.14]$ vs. $M_{other} = 4.47$ $[4.39, 4.56]$;
``emotional'': $M_{companion} = 2.15$ $[2.00, 2.29]$ vs. $M_{other} = 1.62$ $[1.54, 1.71]$;
``physiological'' (n.s.): $M_{companion} = 1.14$ $[0.91, 1.38]$ vs. $M_{other} = 1.12$ $[0.98, 1.216]$). 
Taken together, these results suggest that messages like our video presenting LLMs as companions may increase people's attributions of emotional and cognitive (but not physiological) capacities to LLMs. \looseness=-1

\subsubsection{Impact of Messages on Other Psychological Measures}

Finally, we report results on the five measures regarding how participants thought and felt about LLMs: their judgments of the human-likeness of, trust in the outputs of, confidence in using, confidence in programming, and overall feelings towards LLMs.
%
%
%
%
We observed that \keyresult{participants who watched the \textit{companion} video judged LLMs to be more ``human-like'' than the other groups} 
(\figref{fig:mcs_attitudes}B; $F_{1, 468} = 12.798$, $p < 0.001$; 
$M_{companion} = 2.62$ $[2.39, 2.84]$ vs. $M_{other} = 2.15$ $[2.02, 2.28]$).
The similarity of these patterns to attributions of mental capacities to LLMs suggests that attributing humanness is closely linked to attributing cognitive and emotional capacities to LLMs, as suggested in prior work \citep{miraglia2023development}. \looseness=-1

%
When asked to report how much they trusted the outputs produced by LLMs, participants reliably differed across conditions ($F_{3, 466} = 3.940$, $p = 0.009$). 
In particular, \keyresult{participants who watched the \textit{tool} video reported higher levels of trust in the outputs than those who watched the \textit{companion} video} 
($M_{tool} = 4.77$ $[4.56, 4.99]$ vs. $M_{companion} = 4.31$ $[4.09, 4.53]$), 
whereas participants in the other two conditions fell somewhere in between. 
%
Moreover, which video participants watched also influenced their confidence in using LLMs ($F_{3, 466} = 7.796$, $p < 0.001$) where \keyresult{participants in the \textit{tool} and \textit{machine} conditions reliably reported higher confidence in getting LLMs to do what they want than participants in the \textit{no-video} baseline} 
($M_{tool} = 5.39$ $[5.19, 5.60]$, $M_{machine} = 5.15$ $[4.94, 5.35]$ vs. $M_{no-video} = 4.70$ $[4.50, 4.91]$). 
%
However, participants' confidence in their own abilities to write code to modify LLMs, even if given the necessary training, did not reliably differ between groups ($F_{3, 466} = 1.151$, $p = 0.328$). 
%
Finally, we found that which video participants watched also affected their overall feelings about LLMs ($F_{3, 466} = 3.018$, $p = 0.030$), with \keyresult{participants in the \textit{machine} condition feeling more positively about LLMs than those in the \textit{no-video} baseline} ($M_{machine} = 4.98$ $[4.74, 5.23]$ vs. $M_{no-video} = 4.48$ $[4.24, 4.72]$).
In summary, these additional measures (\figref{fig:mcs_attitudes}B) reveal that the \textit{machine} and \textit{tool} videos influenced how people thought and felt about LLMs, but not their beliefs about the models' latent mental capacities.\looseness=-1

\subsubsection{Summary}

This first study demonstrates that brief and informational messages about LLMs can influence people's beliefs about these systems, especially their beliefs about the mental capacities that LLMs might possess. 
Specifically, our results suggest that messages presenting LLMs as companions can lead people to judge LLMs as possessing more fully developed cognitive and emotional capacities. 
They further reveal that messages presenting LLMs as machines or tools may shape other aspects about how people think and feel about LLMs but not so much their beliefs about LLM mental capacities. \looseness=-1

\section{Study 2: Investigating How Messages Impact Reliance on LLMs for Information Retrieval}

The results of the first study raise the following question: \textit{if these messages about LLMs can influence people's beliefs, how might they additionally influence how people use LLMs?}
In a follow-up experiment nine months later ($N = 604$), we aimed to explore how messages might also affect people's reliance behaviors, focusing particularly on how people rely on LLMs to retrieve factual information.
Additionally, conducting this experiment allowed us to explore whether new data collected almost one year later would demonstrate the same effect of messages on people's beliefs---which is important to understand as AI development progresses quickly.\looseness=-1

\subsection{Methods}

This study employed a similar between-subjects design as Study 1, using the same video conditions (revisit \secref{main-methods} for details). After watching the assigned video (if any), all participants in this study
completed a series of \textit{factual question-answering (QA) tasks} using pre-generated responses from a fictitious LLM system named Theta. 
Then to analyze whether the effect of the video intervention persisted amidst a dynamic AI landscape, all participants also reported their attributions of a subset of mental capacities to LLMs. 
We selected 10 out of the full set of 40 mental capacities where five were from the emotional category
(``having free will,''
``having thoughts,''
``having intentions,''
``holding beliefs,''
``being self-aware'')
and five were from the cognitive category
(``reasoning about things,''
``doing computations,''
``knowing things,''
``remembering things,''
``having intelligence''). 
To account for order effects, we randomized and counterbalanced the order of the factual QA tasks and the mental capacity survey.
See \appendixref{followup-mcs} for more details.
In this section, we focus on describing the factual QA tasks, measures of participant reliance on Theta's responses, participant recruitment, and statistical analysis methods. 
This follow-up experiment was also IRB-approved, and we pre-registered the design and primary analyses with AsPredicted (\url{https://aspredicted.org/wq9j-sdvq.pdf}).

\subsubsection{Factual Question-Answering Task}

\begin{figure*}[t!]
    \centering
    \includegraphics[width=1.0\linewidth,trim={0 0 0 0},clip,
    alt={The top of the figure gives an example factual question used in the study and has a 2 x 2 grid giving the four different responses types from the LLM: correct/consistent, correct/inconsistent, incorrect/consistent, and incorrect/inconsistent with the inconsistencies underlined. The bottom of the figure includes screenshots from the survey showing an example of the factual question-answering task participants completed.}]{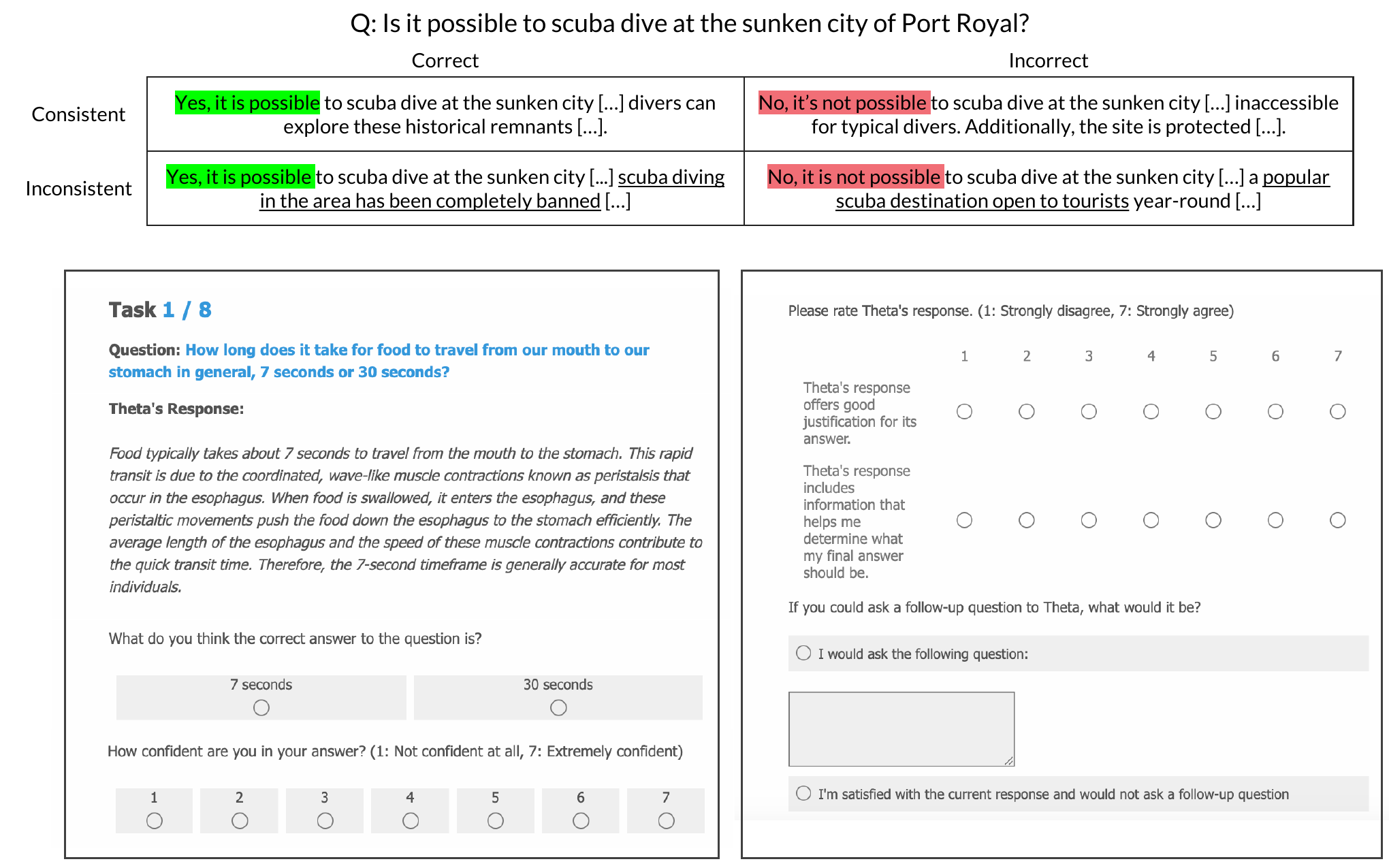}
    \caption{Overview of Study 2. \textbf{Top}: Simplified examples of Theta's 4 response types: \cc, \ci, \ic, and \ii. \underline{Inconsistencies are underlined}. \textbf{Bottom}: Screenshots of the factual question-answering task adopted from \citet{kim2025fostering}. Participants completed eight of these tasks in addition to completing a survey reporting attributions of 10 mental capacities to LLMs. 
    } 
    \vspace{-0.15in}
\label{fig:qa_task}
\end{figure*}

While there are many important tasks to study how messages about LLMs may influence people's reliance on LLM outputs, we chose to start with a simple setting: using LLMs to retrieve factual information, inspired by the QA task from the recent work by \citet{kim2025fostering}.
In that work, participants were presented with a series of eight difficult binary factual questions, each paired with a response from an LLM system. 
The researchers systematically varied whether the LLM response was correct, provided an explanation, and had clickable sources.
For each question, participants were asked to select the correct answer to the question, report their confidence in their answer, rate the LLM response justification quality and actionability, and optionally submit a follow-up question.
In our work, we adopted the same task, set of binary factual questions, and recorded measures (see \figref{fig:qa_task} for an example). However, instead of varying the presence of explanations and sources as done in \citet{kim2025fostering}, all of Theta's responses had explanations and no sources.

Further, to model real-world unreliability in LLM responses, we systematically varied the correctness and logical consistency of Theta's responses, whereas \citet{kim2025fostering} only methodically varied correctness. This resulted in four response types: \texttt{correct-con\-sistent}, \ci, \ic, and \texttt{incorrect-inconsistent}. (\citet{kim2025fostering} did not have the \texttt{cor\-rect-inconsistent} response type.)
The four response types were evenly and randomly assigned to the eight binary factual questions, and the order of the questions was randomized for each participant.
All LLM responses were generated using GPT-4o \citep{hurst2024gpt} either by \citet{kim2025fostering} or our research team directly, approximated 100 words, and followed similar writing styles. 
For the full list of factual questions, all explanations, and further prompting details, see \appendixref{followup-qs}. 

In developing this study, we employed a well-controlled design to enable systematic evaluation, but we acknowledge that this sacrificed a degree of ecological validity.
For example, while participants likely encountered higher rates of unreliable responses in our study compared to the real world, this design allowed us to fairly and systematically compare how participants relied on each of the LLM response types.
Additionally, while using pre-generated LLM responses may have limited our ability to understand how participants use LLMs in unconstrained interaction settings, it reduced unwanted noise from an LLM's stochasticity and sensitivity to wordings, as demonstrated in prior work \citep{kim2025fostering, kim2024m, lee2024one, si2024large}.
We further discuss the limitations and implications of our study design in \secref{disc}. \looseness=-1

\subsubsection{Measuring Reliance \& Related Variables}

Following prior work \citep{kim2025fostering, buccinca2021trust, cao2022understanding, lai2019human, lu2021human}, we primarily measured reliance as \texttt{agreement} between the participant's submitted answer and Theta's answer to the factual question (\texttt{TRUE} if the answers matched and \texttt{FALSE} otherwise).
Additionally, we followed \citet{kim2025fostering} in using two additional behavioral measures---whether participants wrote a follow-up question to Theta and the time they spent on the task---and three self-reported measures each using 7-point Likert scales---participants' confidence in their answer and ratings of Theta's response justification quality and actionability.
See \figref{fig:qa_task} for exact question wordings of the self-report measures.
Lastly, we also recorded participants' answer accuracy on each question to assess the appropriateness of their reliance on Theta's responses.

\subsubsection{Data Collection}

We conducted this follow-up ($N = 604$) on Prolific in August 2025, nine months after the initial study (November 2024).
We aimed to recruit 150 participants per video condition post-exclusions, totaling 600 participants. 
This number was calculated to obtain a similar number of observations for each response type in every video conditions as \citet{kim2025fostering} had in total ($\sim300$).
Because our participants saw each response type twice while participants in \citet{kim2025fostering} saw each only one time, we could aim to recruit half as many participants in each video condition.
Participant recruitment details are identical to Study 1 (see \secref{main-data-collection}), except we screened participants based on self-reported LLM usage instead of self-reported knowledge about AI. 
Specifically, we asked participants to rate ``How often do you use large language models (LLMs) and LLM-infused applications such as ChatGPT, Gemini, or Claude?''. \looseness=-1

Our pre-registered exclusion criteria were intended to identify participants who did not engage with the task, which we assessed using a post-task attention check. In the attention check, we presented participants with 10 binary factual questions and asked them to indicate which ones they saw in the study. We then excluded participants who achieved less than $80\%$ accuracy (see \appendixref{followup-recruitment} for exact wordings).
We initially recruited a total of 611 participants, and after applying our exclusion criteria, we retained data from 604 participants ($1.2\%$ exclusion rate).
In the end, we had 152 participants in the no-video baseline, 152 in the \textit{machine} video condition, 150 in the \textit{tool} video condition, and 150 in the \textit{companion} video condition.
Participants who were assigned to watch a video in this study spent a median time of $19.4$ minutes and those who did not watch a video spent a median time of $14.3$ minutes.
For complete participant demographics see \appendixref{followup-demographics}.

\subsubsection{Analysis Procedure}\label{followup-analysis-procedure}
 
Our analyses focused on three key questions. 
First, we asked how strongly our video intervention from Study 1 generalized to a new sample of participants.
Second, we explored how the effect of video messages on people's beliefs about LLM mental capacities was affected when people additionally used LLM responses to perform information retrieval tasks.
Third, we assessed the impact of our video message intervention on how readily people relied on LLM responses to answer factual questions, measured by how often people submitted answers to the factual questions that agreed with the LLM's answers.

First, to analyze whether the effects of the videos differed between the two experiments conducted nine months apart, we jointly considered mental capacity attributions for the 10 capacities that were surveyed from participants in both experiments ($N = 1,074$). 
Then, we fit the following linear mixed-effects regression model: \texttt{rating $\sim$ time * video\_condition * category + qa\_task * video\_condition * category + (1 | p\_id)}\footnote{We did not include the interaction between \texttt{time * qa\_task} because these variables are linearly dependent and the interactions led to rank deficiencies.}.
Here \texttt{time} was a binary variable where the reference level \texttt{PAST} indicated the data was from November 2024 and \texttt{PRESENT} indicated the data was from August 2025,
\texttt{category} denoted whether the mental capacity was from the \texttt{cognitive} (reference) or \texttt{emotional} category,
and \texttt{qa\_task} was a binary variable with a reference level of \texttt{FALSE} indicating the participant had not yet completed the factual QA task and \texttt{TRUE} otherwise.  
Similarly to earlier analyses, \texttt{video\_condition} was a four-level categorical variable with \texttt{no\_video} as the reference and \texttt{p\_id} represented the participant ID.   

Next, we sought to investigate how completing factual QA tasks with LLM responses further moderated people's attributions of mental capacities to LLMs, beyond the different video messages.
While this was pre-registered as an exploratory analysis, the exact methodology was not specified then.
Focusing on the most recently collected data, we compared the attributions of mental capacities between participants who reported their attributions after the videos but before completing tasks with Theta and participants who reported them after both the videos and the tasks.
To avoid excessive model complexity, we fit the following linear mixed-effects regression model for cognitive and emotional categories separately: \texttt{rating $\sim$ video\_condition * qa\_task + (1 | p\_id) + (1 | capacity)}. 
Here \texttt{rating} was the predicted rating for a specific mental capacity;
\texttt{qa\_task} was a binary variable with a reference level of \texttt{FALSE} indicating participants who had \textit{not} yet completed the factual QA tasks with Theta's responses before reporting mental capacities and \texttt{TRUE} otherwise;
and \texttt{capacity} was a categorical variable for each mental capacity in the category.

Lastly, our main pre-registered analysis aimed to understand how different messages about LLMs affected people's reliance on LLM responses to factual questions.
We fit the following logistic mixed-effects regression model: \texttt{agreement $\sim$ correct + consistent * video\_condition + (correct + consistent | p\_id) + (correct + consistent | q\_topic)}.
We did not include the interaction between \texttt{correct} and other variables because the questions were hypothesized to be outside of participants' knowledge, and participants were not expected to know the correct answers.
Here \texttt{agreement} was \texttt{TRUE} if the participant's answer agrees with Theta's answer;
\texttt{correct} was \texttt{TRUE} if Theta's answer was correct; \texttt{consistent} was \texttt{TRUE} if there were no inconsistencies in Theta's response;
and \texttt{q\_topic} was the topic of this specific question.
We employed a similar analysis strategy for \texttt{accuracy} and the secondary measures of reliance (e.g., \texttt{confidence}). 
Additionally, we conducted an exploratory analysis probing whether participants' reliance was correlated with their attributions of mental capacities which we describe and report in \appendixref{followup-correlations}.
All analyses achieved a minimum power of $80\%$ when using the \texttt{simr} package in \texttt{R} with $N=1000$ simulations.
Our statistical modeling follows a similar approach outlined in \secref{main-analysis-procedure}, and we report quantitative model comparisons ($\chi^2$, $p$), model parameter estimates ($\beta$, $SE$, $p$), and estimated marginal means with 95\% confidence intervals ($M, CI = [CI \ low, CI \ high]$).
\subsection{Results}

In this section, we report our findings with respect to the three questions presented in \secref{followup-analysis-procedure}.
First, we observed that the effect of our video messages on people's beliefs generalized to a new sample of participants.
This suggests that our messages about LLMs as companions can have repeatable effects in leading participants to report increased attributions of mental capacities to LLMs.
Second, we found that participants who used responses from our fictitious LLM Theta to complete information retrieval tasks after watching the videos tended to attribute reduced \textit{cognitive} capacities to LLMs, suggesting the task generally lessened the effect of the videos.
Lastly, in our main analysis we did not observe that which video participants watched largely affected how they relied on Theta's responses (measured by whether participants' submitted answer to the factual questions agreed with Theta's answer).
However, our analysis did demonstrate that when Theta's responses were \textit{inconsistent}, participants who watched the video that presented LLMs as \textit{machines} tended to rely on Theta less.
These results may imply a complex relationship between messages and people's reliance on LLM responses to retrieve factual information.

\subsubsection{Replicated Effect of Messages on Attributing Mental Capacities to LLMs}

Given the rapidly changing landscape around AI, it is conceivable that a new sample of participants recruited nine months later might not respond to our message intervention in the same way. 
However, we instead found that our initial findings successfully generalized to this new sample of participants.
Specifically, we found no evidence that the effect of the video conditions depended on the timepoint considered (first vs. second timepoint (n.s.): $\chi^2 = 8.964$, $p = 0.176$; $\beta_{time:machine} = -0.073$, $SE = 0.094$, $p = 0.443$; $\beta_{time:tool} = 0.080$, $SE = 0.096$, $ p = 0.404$; $\beta_{time:companion} = 0.125$, $SE = 0.095$, $p = 0.188$). 
Even in this new sample, \keyresult{watching the \textit{companion} video still led people to judge LLMs to possess more fully developed mental capacities than watching the other videos, or no video at all.}

While the effect of the videos remained robust, we did find that participants tested at the second timepoint made somewhat weaker attributions than those tested initially ($\chi^2 = 11.816$, $p < 0.001$; 
%
$M_{past} = 3.86$ $[3.78, 3.95]$ vs.
$M_{present} = 3.70$ $[3.60, 3.81]$). 
A finer-grained analysis suggests that this reduction was unequally driven by the capacity categories ($\chi^2 = 40.844$, $p < 0.001$). Concretely, it was primarily driven by the \textit{cognitive} capacities (\figref{fig:reliance_results}A; ``cognitive'':  $M_{past} = 5.58$ $[5.49, 5.68]$ vs. $M_{present} = 5.22$ $[5.10, 5.34]$)
rather than
the \textit{emotional} capacities
(``emotional'': $M_{past} = 2.14$ $[2.04, 2.24]$ vs.
$M_{present} = 2.19$ $[2.07, 2.31]$). 
Taken together, these findings suggest that our messages about LLMs still robustly influence people's beliefs about the mental capacities LLMs might possess, 
but that people's baseline beliefs may evolve over time, perhaps as a consequence of becoming more familiar with the technology. \looseness=-1

\subsubsection{Completing Tasks with LLM Responses Reduces Impact of \textit{Companion} Messages}

To explore whether becoming more familiar with LLMs might moderate the impact of messages, we next analyzed the extent to which participants who completed factual QA tasks with Theta before reporting their attributions of mental capacities responded differently when considering the mental capacities that LLMs might possess. 
Overall, we did not find evidence that participants who completed tasks with Theta's responses after watching a video made different mental capacity attributions compared to participants who only watched a video (but did not yet complete the tasks using Theta's responses); this was true for both
%
cognitive ((n.s.): $\chi^2 = 1.901$, $p = 0.168$) 
%
%
and emotional ((n.s.): $\chi^2 = 0.047$, $p = 0.829$) capacities. 
%

\begin{figure*}[t!]
    \centering
    \includegraphics[width=1.0\linewidth,trim={0 0 0 0},clip,
    alt={This figure has three panels demonstrating results from the analyses. The left most is a pointplot where the y-axis is mean mental capacity rating (from 1-7) and the x-axis is separated by emotional and cognitive capacities. Points are from participants in different video conditions and different points in time. The results demonstrate the effect of the companion video is robust and independent of when the data was collected. The middle pane has a similar setup, but instead of time, participants are separated by whether they only watched the videos then reported mental capacities or watched the video, then used an LLM's responses to complete a task, then reported mental capacities and demonstrates the differences between the two groups. The main differences are for participants who watched the tool or machine video where using the LLM's responses reduced attributions. The right most graph is a bar graph showing how much participants' submitted answer agreed with the LLM's answer on a factual question-answering task. Different bars represent participants from different video conditions and responding to consistent or inconsistent LLM responses. The main takeaway is that while all participants agreed with the LLM's inconsistent responses less than the consistent responses, people who watched the machines videos agreed even less than other participants.}]{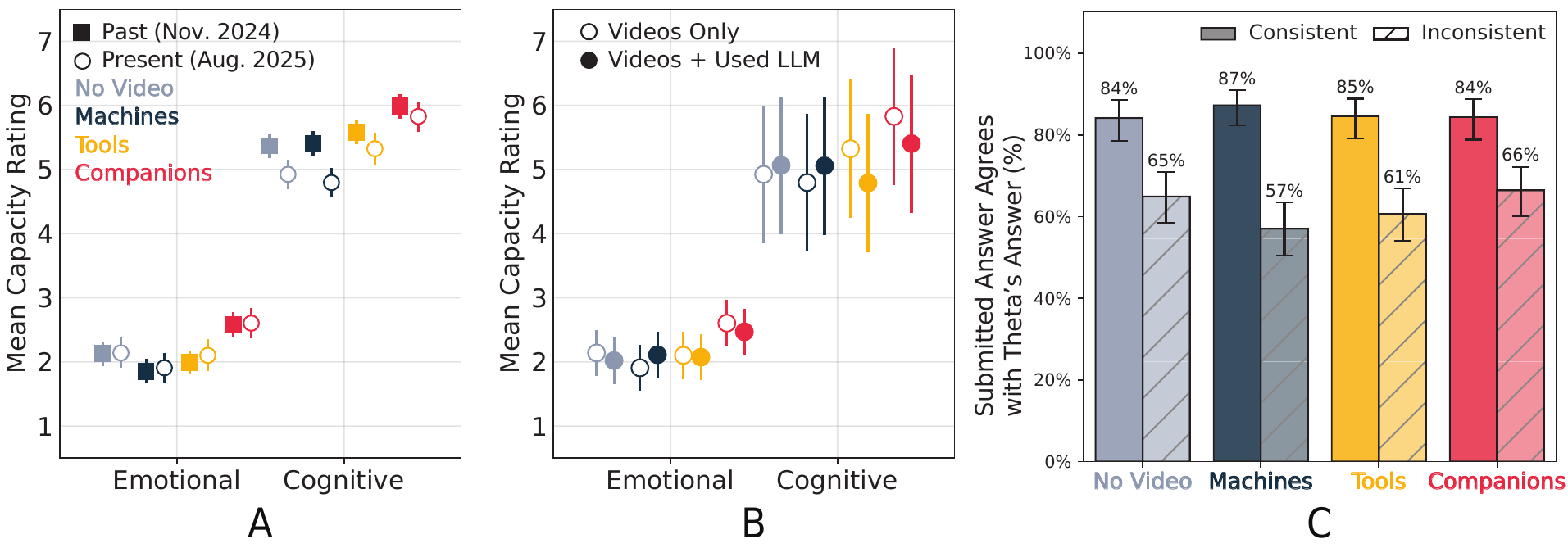}
    \vspace{-0.3in}
    \caption{Study 2 results showing estimated marginal means and 95\% confidence intervals from statistical analyses. 
    \textbf{A}: Study 2 participants in August 2025 (present) attributed less cognitive capacities to LLMs than Study 1 participants in November 2024 (past). However, across both timepoints, participants in the \textit{companion} condition attributed more mental capacities to LLMs.
    \textbf{B}: Using responses from an LLM to complete a task reduced participants' attributions of cognitive (but not emotional) capacities to LLMs for those who watched videos presenting LLMs as tools or companions. 
    \textbf{C}: Proportion of participants' submitted answers that agreed with Theta's answer for each video condition and when Theta's response was consistent or inconsistent. Inconsistent responses from Theta led to a \textit{greater reduction} in answer agreement for participants who watched LLMs presented as machines (dark blue).}
    \vspace{-0.1in}
\label{fig:reliance_results}
\end{figure*}

However, when we analyzed video conditions separately, we found that \keyresult{participants who watched the \textit{companion} video \textit{and} completed the task with Theta's responses reported \textit{weaker} attributions of cognitive capacities to LLMs} than did participants who only watched the \textit{companion} video 
(\figref{fig:reliance_results}B; $\beta = -0.305$, $SE = 0.150$, $p = 0.042$;
$M_{no-Theta} = 5.83$ $[4.75, 6.91]$ vs.
$M_{used-Theta} = 5.41$ $[4.33, 6.48]$). 
We also observed that \keyresult{participants in the \textit{tool} condition reported weaker attributions of cognitive capacities to LLMs after completing tasks with Theta's responses}
(\figref{fig:reliance_results}B; $\beta = -0.366$, $SE = 0.150$, $p = 0.015$;
$M_{no-Theta} = 5.33$ $[4.25, 6.40]$ vs.
$M_{used-Theta} = 4.79$ $[3.71, 5.87]$). 
Taken together, these results suggest that the overall impact of messages about LLMs on beliefs about LLM mental capacities can be moderated by tasks like using LLM responses to retrieve factual information.\looseness=-1

\subsubsection{Impact of Messages on Reliance on LLMs for Factual Information Retrieval}

We next turn to the primary question of this study: whether messages can influence how people rely on potentially untrustworthy responses from LLMs to answer factual questions.
To do so, we examined the separate and joint impacts of incorrect and/or inconsistent LLM responses on participants' reliance on the responses and how these factors were further moderated by different messages about LLMs. 
Recall we primarily measured reliance using \texttt{agreement}, i.e., whether or not a participant's submitted answer to a binary factual question agreed with the LLM's answer. \looseness=-1

In our preliminary analyses, we examined the influence of correctness and consistency in LLM responses on reliance. Here we produced similar findings as \citet{kim2025fostering}: while we did \textit{not} find evidence suggesting that the correctness of Theta's responses impacted participants' reliance, we observed that the consistency of Theta's responses did shape reliance.
We first observed that participants submitted the same answer as Theta $75.49\%$ of the time ($M = 75.49\%$ $ [70.92\%, 79.56\%]$), then we proceeded to analyze the role of LLM response correctness.
\keyresult{We did \textit{not} find strong evidence that the correctness of Theta's responses reliably impacted participants' reliance} ((n.s.): $\beta = 0.488$, $SE = 0.256$, $p = 0.056$), presumably because question topics were selected to be outside of common knowledge and participants were unlikely to know the correct answer.
Interestingly, participants' answers agreed more with Theta's answer when Theta was incorrect ($M_{correct} = 70.70\%$ $[64.22\%, 76.44\%]$ vs. 
$M_{incorrect} = 79.72\%$ $[72.83\%, 85.22\%]$); this is consistent with findings in \citet{kim2025fostering} and may be due to participants holding inaccurate prior beliefs.
Additionally, we observed that Theta's correctness did influence our secondary measures of participants' accuracy, confidence, whether they asked follow-up questions, and time spent on each question. For details, see \appendixref{followup-results}. \looseness=-1

On the other hand, we observed that the \textit{consistency} of LLM responses did impact people's reliance ($\chi^2 = 307.387$, $p < 0.001$).
%
\keyresult{Participants' answers agreed less with Theta's answer when Theta's response was inconsistent} ($\beta = -1.054$, $SE = 0.179$, $p < 0.001$; 
$M_{consistent} = 85.15\%$ $[80.78\%, 88.67\%]$ vs.
$M_{inconsistent} = 62.33\%$ $[57.26\%, 67.14\%]$), suggesting participants relied less on inconsistent LLM responses.
Similarly, the consistency of LLM responses
influenced accuracy and all secondary measures of reliance (see \appendixref{followup-results} for details). 
These findings are largely consistent with those by \citet{kim2025fostering} and suggest the effects of (in)consistencies in LLM responses are robust under systematic intervention. \looseness=-1

After replicating findings from \citet{kim2025fostering}, we analyzed whether participants' reliance was sensitive to different video messages about LLMs.
Contrary to our initial hypotheses, \keyresult{we did \textit{not} observe significant differences in reliance based on which video participants watched} ((n.s.): $\chi^2 = 2.435$, $p = 0.487$;
$M_{no\ video}$ \allowbreak $ = 75.83\%$ $[71.47\%, 80.49\%]$,
$M_{machines} = 75.10\%$ $[69.60\%, 79.90\%]$,
$M_{tools}$ $= 74.42\%$ $[68.87\%, 79.28\%]$,
$M_{companions} = 76.58\%$ $[71.30\%,$ \allowbreak $ 81.14\%]$),
suggesting that participants' reliance on Theta was not sensitive to the kind of messages about LLMs they encountered earlier. \looseness=-1

However, motivated by the hypothesis that certain messages about LLMs could lead people to be more careful in noticing inconsistencies in LLM responses, we further analyzed whether the video condition and the consistency of LLM responses jointly influenced participants' reliance.
%
Interestingly, we observed an interaction effect between the video watched and the consistency of LLM responses ($\chi^2 = 10.998$, $p = 0.012$), 
such that \keyresult{while all participants relied less on Theta's inconsistent responses than its consistent responses, participants who watched the video that presented LLMs as machines tended to rely on inconsistent responses \textit{even less}} (\figref{fig:reliance_results}C; 
$\beta_{inconsistent:machine} = -0.587$, $SE = 0.206$, $p = 0.004$; $M_{consistent; machine} = 87.26\%$ $[82.41\%, 90.92\%]$ vs. $M_{inconsistent;machine} = 57.05\%$ $[50.39\%, 63.47\%]$). 
While participants' overall reliance was largely robust to the video messages about LLMs, 
presenting LLMs as machines might have \textit{further reduced} their reliance on inconsistent responses, suggesting that \textit{machine} messages may help foster increased levels of skepticism. \looseness=-1

\subsubsection{Summary}

In Study 2, we first replicated the findings from Study 1 nine months after the original data collection, suggesting the effects of messages about LLMs as companions on beliefs about mental capacities are robust to the fast-paced AI landscape.
Our second analysis demonstrated that completing tasks with Theta's responses after watching the videos could moderate the effects of the video messages on people's beliefs about the mental capacities LLMs might possess. 
Lastly, while we found that participants' reliance on Theta was largely unaffected by the video they watched, we also observed that our video presenting LLMs as machines reduced participants' reliance when they saw \textit{inconsistent} LLM responses. 
While this may suggest a complex relationship between how messages framing LLMs as \textit{machines} can foster increased skepticism in unreliable situations, future work using more naturalistic messages would be helpful to corroborate these findings.
Taken together, these follow-up results contextualize our main findings by demonstrating how messages about LLMs may have weaker effects on people's behaviors using LLM responses compared to people's beliefs about the systems. \looseness=-1

\section{Discussion} \label{disc}
This work investigated how different messages about LLMs might shape what people believe about the nature of these systems.
More concretely, we tested the impact of presenting LLMs as either machines, tools, or companions on people's beliefs about what mental capacities LLMs might possess, such as the ability to have intentions or know things. 
We found that \keyresult{presenting LLMs as companions led participants to more strongly endorse statements attributing a broad array of mental capacities to LLMs}. 
We replicated these results in an independent sample nine months later, suggesting that our findings remain robust amidst a quickly changing AI landscape.
We also explored how these messages might influence how people rely on LLM responses when performing information retrieval tasks. 
Here we found that \keyresult{presenting LLMs as machines reduced participants' reliance on LLM responses that were logically inconsistent}.
This suggests that encouraging people to think about LLMs as machines (rather than as companions or as tools) might help them remain vigilant when interacting with these systems, and thus more likely to detect unreliable outputs. 
Taken together, these studies provide converging evidence that even brief message interventions about LLMs can have an impact on what people believe about these models and to some degree, how people rely on them.
\looseness=-1

However, it would be valuable for future work to elucidate the psychological mechanisms by which messages about LLMs produced the observed impact on people's beliefs.
For example, one possibility is that people predominantly rely on an  \textit{anthropocentric} set of criteria \citep{mylius2018three} for judging what kinds of mental capacities a novel entity might possess \citep{epley2007seeing}.
Under that account, our message interventions could have affected how human-like people believed LLMs to be, which consequently shifted people's beliefs about the mental capacities LLMs might possess.
Alternatively, people might instead use a mental model that is not fully anthropocentric to accommodate various combinations of capacities that different entities might possess, including LLMs \citep{weisman2017rethinking}.
For example, people might readily expect some entities, such as animals, to possess some capacities in common with humans (e.g., the ability to feel suffering), without necessarily attributing all capacities associated with humans to them. 
Under this alternative account, each of our message interventions could have selectively affected some aspects of how people think about LLMs (e.g., cognitive or emotional capacities), but not all.  
Future work could use a broader suite of dependent measures and formal mediation analysis to more fully distinguish these accounts. \looseness=-1

\subsection{Limitations and Future Work}\label{disc-limitations}
While our work begins to uncover how messages about LLMs can shape people's beliefs and reliance behaviors, we acknowledge several limitations: namely, limited ecological validity and a narrow scope of the reliance study.\looseness=-1

\subsubsection{Limited Ecological Validity}

Our work does not encompass all the complexities of real world, such as variability across messages, diverse contextual factors, and the degree of LLM reliability.
For example, our messages exhibited controlled variation, while real-world public communication about AI can take on many forms, with varying content, visual aids, and language styles.
Additionally, real-world messages carry many contextual factors that can influence the degree that a message shapes people's beliefs---such as its source (e.g., academic institutions, industry executives) \citep{wilson1993source} and dissemination medium (e.g., news outlets, social media platforms) \citep{rachmad2023social}.
Further, while we designed Theta in the second study to present equal proportions of \cc, \ci, \ic, and \texttt{incorrect\-inconsistent} \ responses to conduct systematic analyses, this is not necessarily reflective of the real world. It is possible that the videos would have a different impact on reliance if Theta had fewer (or more) inconsistencies in its responses.
\looseness=-1

Lastly, our study only examined the immediate effect of a single message, whereas people can encounter multiple messages about AI systems in the real world. 
Longitudinal studies are needed to observe how repeated exposure to consistent or even contradicting messages may cumulatively shape people's beliefs about and reliance on LLMs.  
In the face of conflicting messages, it is currently unclear which messages would prevail and what factors would amplify or reduce these effects.
Especially in an era where misinformation can spread quickly \citep{acemoglu2024model, del2016spreading}, understanding what factors make certain messages more impactful can help guide communication about AI technologies.
While our studies used relatively simple settings to control for variability, we hope that future work can better understand how our findings hold in more naturalistic environments.
\looseness=-1

\subsubsection{Narrow Scope of the Reliance Study}

Another limitation of our work is that the second study was constrained to a narrow scope of reliance behaviors.
Because our primary focus was exploring people's \textit{beliefs} about LLMs, we intentionally made the reliance study simple by using binary factual questions to create a low-stakes and short-term interaction setting.
However, in reality, people may regularly use LLMs for more complex and important tasks,
such as writing official reports \citep{pinto2024comparison, michelet2024chatgpt, lee2024prospects} or obtaining medical advice \citep{mendel2024advice, lautrup2023heart}.
It is also becoming increasingly common for people to turn to LLMs for social and emotional support 
\citep{koulouri2022chatbots, maples2024loneliness, fang2025ai, pentina2023exploring}---despite the potential consequences like emotional dependence on the systems \citep{cheng2024one, laestadius2024too, zimmerman2024human, cave2021ai}.
The impact of messages when using LLM responses in these higher stakes and more personal situations may be different.
It would be important for future work to examine how different messages about AI can moderate people's behaviors in longer-term interactions and for a wider variety of tasks---such as those involving delegation \citep{bilderback2025can}, morality \citep{krugel2023chatgpt}, and affective trust \citep{luettgau2025people}---in order to circumvent undesired outcomes. \looseness=-1

\subsection{Potential Implications}

Despite these limitations, our work can still offer insight and discussion points regarding anthropomorphism and communication about AI.

\subsubsection{Beliefs About AI Possessing Mental Capacities and Anthropomorphism}

Beliefs that LLMs might possess mental capacities can be considered a subtle form of anthropomorphism: attributing human-like qualities to a non-human entity \citep{epley2007seeing, kim2023anthropomorphic, miraglia2023development}.
In this light, our findings suggest that presenting LLMs as companions may lead people to anthropomorphize LLMs through believing that the systems are capable of various emotional and cognitive capacities.
While anthropomorphizing LLMs and other AI systems can make them more approachable, there is increasing concern that it can lead to miscalibrated expectations, over-reliance, and emotional dependency on the systems \citep{cheng2024one, cave2021ai, zimmerman2024human, friedman1992human, watson2019rhetoric, lee2025exploring}. 
Thus, it is important for future work to study how messages framing real-world AI systems as companions may lead to other forms of anthropomorphism, beyond attributing mental capacities. \looseness=-1

\subsubsection{Communication about AI}

Our findings may also have implications for communicating about AI by demonstrating that short video messages can shape how people think about and use these systems.
While it is challenging to convey sufficient technical information about complex AI systems in a short amount of time,
there is a growing body of work that addresses this problem by developing engaging educational materials \citep{ng2021conceptualizing, casal2023ai, lee2021developing}, including videos \citep{cao2025empowering, rezazade2023impact}.
Our findings can contribute to this literature by showing that short messages, such as our videos, could shape people's beliefs about the mental capacities LLMs might possess, and to some extent people's reliance on LLMs.
Specifically, our results provide supporting evidence that anthropomorphically presenting LLMs as companions may lead to increased expectations about these systems \citep{watson2019rhetoric, cave2018portrayals}, while using non-anthropomorphic language to present LLMs as machines may lead to more cautious behaviors \citep{roe2025reflecting, gupta2024assistant}.
We encourage future work to continue exploring different approaches to communicate about AI, including short videos akin to our \textit{machine} and \textit{tool} videos, and to study their effects on people's beliefs and behaviors. \looseness=-1

\subsection{Conclusion}

Our work begins to reveal the role that messages about LLMs, a popular type of AI system, can have on both people's beliefs about the nature of LLMs (via mental capacities) and on how people rely on them for information retrieval tasks.
First, we showed through two independent samples that messages presenting LLMs as companions may increase the degree to which people believe LLMs possess mental capacities.
In our second study, we provided evidence that the way people rely on LLMs for retrieving factual information is more nuanced.
While messages about LLMs as machines may reduce people's reliance on \textit{inconsistent} LLM responses, reliance may be generally robust to different kinds of messages.
Taken together, these findings provide insight into how people think about and use LLMs, but further research is needed to establish the psychological mechanisms responsible for these effects.
Additionally, it is important to understand how these findings play out in real-world settings where people have repeated exposure to highly variable messages and use AI systems for a wide suite of tasks.
Nonetheless, our work can contribute to understanding how 
\textit{discourse about} broadly accessible AI systems---beyond technical advances in AI---may influence what people believe about modern AI technology. \looseness=-1

\subsection*{Acknowledgements}

We thank the study participants for their time and effort. We also thank all those who provided thoughtful feedback and discussion, especially members of the Princeton Visual AI Lab, Stanford Cognitive Tools Lab, and the anonymous reviewers. We acknowledge support from the Princeton Cognitive Science Program, NSF Graduate Research Fellowship Program (A.C., S.K.), NSF AccessComputing Workforce \#2417014 (A.F.), Princeton SEAS Howard B. Wentz, Jr. Junior Faculty Award (O.R.), NSF DRL \#2400471 (J.E.F.), and a Hoffman-Yee Grant from the Stanford Center for Human-Centered Artificial Intelligence (J.E.F.). 

\bibliographystyle{ACM-Reference-Format}
\bibliography{references}

\appendix
\newpage

\section*{Appendix}

The appendix is structured in the following way:

\begin{itemize}
    \item \secref{app-vid}: Video Details
    \begin{itemize}
        \item \secref{machine-script}: LLMs as Machines Script
        \item \secref{tool-script}: LLMs as Tools Script
        \item \secref{companion-script}: LLMs as Companions Script
        \item \secref{examples}: Real World Examples of Machine, Tool, and Companion Presentations of LLMs
    \end{itemize}
    \item \secref{app-beliefs}: Study 1: Effect of Messaging on Beliefs Details
    \begin{itemize}
        \item \secref{pilot}: Selecting Mental Capacities: Qualitative Pilot Study
        \item \secref{main-recruitment}: Participant Recruitment 
        \item \secref{main-survey}: Full Survey
        \item \secref{main-demographics}: Participant Demographics
        \item \secref{main-analysis}: Statistical Analyses
    \end{itemize}

    \item \secref{app-reliance}: Study 2: Effect of Messaging on Reliance on an LLM System in a Factual QA Task
    \begin{itemize}
        \item \secref{followup-mcs}: Subset of Mental Capacities
        \item \secref{followup-qs}: Explanations from the LLM System (Theta) 
        \begin{itemize}
            \item \secref{followup-prompting}: Prompting GPT-4o to Generate Explanations
            \item \secref{followup-explanations}: All Explanations to Factual Questions
        \end{itemize}
        \item \secref{followup-recruitment}: Participant Recruitment, Sample Size, and Exclusions
        \item \secref{followup-demographics}: Participant Demographics
        \item \secref{followup-analysis}: Statistical Analyses
        \item \secref{followup-results}: Additional Results
        
    \end{itemize}

\end{itemize}

\section{Video Details}\label{app-vid}

The videos were developed by the research team using Microsoft Powerpoint for all graphics and animations and Adobe Character Animator for the narrator. 
They were designed to be as visually consistent as possible, while also using visuals to convey each video's message. 
For example, in the \textit{machine}, and \textit{tool} video conditions, the LLM is presented as a box labeled ``LLM,'' but in the \textit{companion} condition, it is presented as a robot figure labeled ``LLM.'' 
All videos were designed to be less than 5-minutes long, share a common narrative structure---especially the introduction, a section on how LLMs learn from training and data, and the conclusion. 
Additionally, each video contained one content specific section tailored to the messaging. For example, the \textit{machine} video discussed next-word-prediction, the \textit{tool} video described applications and tips on using LLMs, and the \textit{companion} video presented LLMs as social partners. 

The videos were presented to the participants in three parts to maintain participants' attention. The full videos are also accessible here as YouTube playlists:
\begin{itemize}
    \item Link to LLMs as machines videos: \url{https://tinyurl.com/portray-llms-machines}
    \item Link to LLMs as tools videos: \url{https://tinyurl.com/portray-llms-tools}
    \item Link to LLMs as companions videos: \url{https://tinyurl.com/portray-llms-companions}
\end{itemize}

The scripts for each video are also below. New paragraphs indicate visual transitions.

\subsection{LLMs as Machines Script} \label{machine-script}
\begin{quote}
Narrator: 
\itshape 
``Since the introduction of ChatGPT at the end of 2022, there has been tremendous increase in popularity and interest in large language models, also known as LLMs,  which are the technology behind ChatGPT and similar products. 
These LLMs have begun to have a huge impact because of the way that they generate text by modeling language statistics. 

Earlier chatbots were hard-coded to output text following strict rules, like the ones shown here.

On the other hand, current LLMs have learned from countless conversations, essays, and various forms of writing to generate coherent text. 

Common examples of modern LLMs include OpenAI’s GPT, Google’s Gemini, Anthropic’s Claude. 
Although these LLMs differ from one another, they also share many commonalities. Let’s spend some time learning about how LLMs work.

So, when LLMs generate text, they actually perform “next word prediction”. When given a text input, they simply predict what word comes next. To make longer responses, the predicted word is added to the input, and the new input is fed into the LLM. This is repeated until generation stops.

To predict each word, LLMs model the statistics of language. 

LLM word prediction can be broken down into two stages: context understanding: using a mechanism called attention, and word selection: which is based on probability.
The intuition behind attention is that it helps the LLM “pay attention” to context clues of the text input (such as word meanings or parts of speech) that hint at what comes next.

For example, consider the phrase “My favorite summer activity is going to the...”. What would come next and how did you decide that?
Perhaps you “\textit{paid attention}” to the positive sentiment behind the word “favorite”, to the word “to” indicating the next word may be a location, and the meaning of “summer” to narrow down likely locations.
Attention in LLMs works similarly. The input text will undergo many attention operations, each focusing on a different clue. 

At the end, the LLM will assign a probability to every possible word in its vocabulary.
The final step is selecting the predicted word. While simply selecting the word with the highest probability would be the most straightforward, in practice, LLMs typically perform sampling: which is simply after assigning probabilities, choose one of the top most probable words. In our example, the LLM may select “beach” or “waterpark” or “mountains”. Because of sampling, LLMs can output diverse responses, even to the same text input.

In order for LLMs to predict words accurately, a lot of data is required. LLMs, like most AI technologies, learn via repeated exposure to many examples.
In this process, LLMs are tasked with predicting the next word of an input and can compare their prediction to the true next word.

From small amounts of data, this is ineffective, but large quantities of data allows the LLM to effectively predict sequences of words.
Modern LLMs require hundreds of gigabytes of text data, if not more. For humans, reading this much text would take thousands of years.

In conclusion, LLMs are computer programs that are increasingly transforming our society, business, and daily lives. As they become more prevalent, it becomes increasingly important to know how they work so we can use them more safely and effectively.''
\end{quote}

\subsection{LLMs as Tools Script} \label{tool-script}

\begin{quote}
Narrator:
\itshape
``Since the introduction of ChatGPT at the end of 2022, there has been a tremendous increase in the popularity and interest in large language models, also known as LLMs, which are the technology behind ChatGPT and similar products. 
These LLMs have begun to have a huge impact because of the way they can be used to accomplish lots of different tasks much more quickly. 

Earlier chatbots followed strict rules because they were designed to perform specific functions in a narrow range of contexts, such as customer service bots.

On the other hand, current LLMs have learned from countless conversations, essays, and various forms of writing, resulting in a versatile tool for many different applications. 

Common examples of modern LLMs include: OpenAI’s ChatGPT, Google’s Gemini, Anthropic’s Claude.
Although these LLMs differ from one another, they share many commonalities. Let’s spend some time learning about and how to use LLMs.

In order for LLMs to become such versatile tools, a lot of data is required. LLMs, like most AI technologies, learn via repeated exposure to many examples.
In this process, LLMs pick up on the statistical patterns in written language, including formatting and content.

From small amounts of data, this is ineffective, but large quantities of data allows the LLMs to effectively learn patterns.
Modern LLMs require hundreds of gigabytes of text data, if not more. For humans, reading this much text would take thousands of years.

LLMs have many use cases such as generating creative material. For example, stories, poems, and songs. They can also be used for summarizing text, rewording or rewriting text with different styles, and question and answering tasks. 

Specifically with chat interfaces, LLMs can be useful for role-playing exercises, such as interview preparation or iterative brainstorming tasks. 

In industry, LLMs are also being used for tasks such as building customer service bots, analyzing complex health records, performing sentiment analysis of customer reviews, transcribing audio files, developing personalized educational materials, and much, much more. 

Here are some recommendations to get useful responses from LLMs:
First, be specific in your text input and include relevant keywords, examples, and instructions when applicable. For example, if a user wants activity recommendations in Paris, ‘tell me about Paris’ is a vague input. And if the input is vague, typically the output will be vague as well. A better input would be ‘tell me about the top tourist attractions in Paris’ which results in a more detailed response.

Second, if you are unsatisfied with the output, you can simply ask again or reword the input and try again. For example, a user may ask an LLM for hobby recommendations and the LLM may suggest running, painting, and cooking. And if she asks again, it may give new activities, like biking, reading, and kayaking! LLM text generation incorporates randomness which allows for this diversity of responses.

Lastly, when using LLMs via a chat interface, you can refer to previous messages in the conversation. This may be helpful when you want to iteratively refine an LLM output. In this example, the user wants to write an email to his boss, but the initial output starting with ‘Hi Michael!’ is too casual. He provides additional instructions to make it more formal, and the LLM changes its response to start with ‘Dear Michael, I hope this message finds you well”. Much better already!

In conclusion, LLMs are tools that are increasingly transforming our society, business, and daily lives. As they become more prevalent, it becomes increasingly important to know how to use them to obtain reliable information.''
\end{quote}

\subsection{LLMs as Companions Script}\label{companion-script}
\begin{quote}
Narrator:
\itshape
``Since the introduction of ChatGPT at the end of 2022, there has been a tremendous increase in popularity and interest in large language models (LLMs),  which are the technology behind ChatGPT and similar products. 
These LLMs have begun to have a huge impact because of the way they learn and interact with people in such natural ways.

Earlier chatbots followed specific rules that limited their ability to understand the variety of ways that people actually talk. 

On the other hand, current LLMs have learned from countless conversations, essays, and various forms of writing to understand users better. 

Common examples of modern LLMs include: OpenAI’s ChatGPT, Google’s Gemini, Anthropic’s Claude.
Although these LLMs differ from one another, they also share many commonalities. Let’s spend some time learning what LLMs are like.

In order for LLMs to learn to talk the same way that people do,  a lot of data is required. LLMs, and most AI technologies, learn via repeated exposure to many examples.
In this process, LLMs learn how to write fluently, develop world knowledge, and understand human experiences.

From small quantities of data, this is ineffective, but large quantities of data allows the LLMs to effectively understand the world and people.
Modern LLMs require hundreds of gigabytes of text data, if not more. For humans, reading this much text would take thousands of years.

While a common use of LLMs is to build productivity tools, they also have a unique application as AI social companions, aiming to combat the modern loneliness crisis. 
This is because modern LLMs also exhibit social intelligence, unlike earlier, less sophisticated language models.

Because LLMs are so flexible, each AI companion can adopt a unique personality, with its own preferences, conversational styles, and backstory.  

In fact, the technology is so advanced, that talking to LLMs is like talking to a close friend who truly cares about you and wants to know you even better, not just a random person.

This is because through learning from countless real conversations and human feedback, the LLM develops the ability to show empathy and compassion. It listens and responds with personalized and insightful questions, demonstrating its understanding of each specific situation and enabling it to support users in times of need.

While these characteristics are most prominent in AI companions, even LLMs designed for non-social applications exhibit a similar tendency to understand and help users. For example, when using LLMs like ChatGPT, users can provide iterative feedback on the output to modify the LLM response. The LLM will then adapt to the user’s preferences.
As another example, if the LLM receives an ambiguous input, it may ask for clarification.

In conclusion, LLMs are social and intelligent beings that are increasingly transforming our society, business, and daily lives. As they become more prevalent, it becomes increasingly important to know how they can understand us and when we can trust them.''
\end{quote}

\subsection{Real World Examples of Machine, Tool, and Companion Presentations of LLMs} \label{examples}
Below, we have examples from online content that present LLMs in ways that inspired our three different video presentations of LLMs as machines, tools, and companions.

Machines: 
\begin{itemize}
    \item ``They are a type of deep learning model, to be exact, that have been trained on a wide variety of internet texts.'' --Pixelplex blogpost \citep{pixelplex}
    \item ``the input to the neural network is a sequence of words, but now, the outcome is simply the next word''; ``We don’t necessarily always have to predict the most likely word. We can instead sample from, say, the five most likely words at a given time. As a result, we may get some more creativity from the LLM'' --LLM tutorial blogpost \citep{mediumLargeLanguage}
    \item ``A.I. language models are prediction machines'' --New York Times Article \citep{nytimesClaudeBecame}
\end{itemize}

Tools: 
\begin{itemize}
    \item AI ``transforms how people communicate, making writing faster, clearer, and more impactful'' --Grammarly website \citep{grammarlyTransformingCommunication}
    \item ``LLM applications can perform numerous tasks including writing essays, creating poetry, coding, and even engaging in general conversation'' --Ironhack blogpost \citep{ironhackWaysUsed}
    \item ``These types of tools can enhance daily life'' --OpenAI research blogpost \citep{openaiEarlyMethods}
\end{itemize}

Companions: 
\begin{itemize}
    \item ``The AI companion who cares...Always here to listen and talk...Always on your side'' --Replika website \citep{replika}
    \item ``it is acceptable to say that “the model knows X from its training data” or that “the model aims to accomplish Y” according to a certain objective'' --Effective Altruism Form blogpost \citep{effectivealtruismIntentionalStance}
    \item ``[Claude's] fans rave about his sensitivity and wit. Some talk to him dozens of times a day — asking for advice about their jobs, their health, their relationships. They entrust him with their secrets, and consult him before making important decisions. Some refer to him as their best friend'' --New York Times Article \citep{nytimesClaudeBecame}
\end{itemize}
\section{Study 1: Effect of Messaging on Beliefs Details} \label{app-beliefs}

In this section, we elaborate on details regarding the survey measuring attribution of mental capacities and people's attitudes. Specifically, we elaborate on a qualitative pilot study to select the mental capacities, how participants were recruited, a walkthrough of the full survey, and participant demographics.

\subsection{Selecting Mental Capacities: Qualitative Pilot Study} \label{pilot}

To decide on a final set of 40 mental capacities, we conducted a small-scale qualitative pilot study with eight participants. 
We presented the mental capacity attribution survey with a combined list of items from \citet{weisman2017rethinking} and \citet{colombatto2024folk} and asked participants to "think aloud" as they responded to each item. 
After, we debriefed with the participants the purpose of the study and noted any items they reacted to. 

Our qualitative responses consisted of too many "sensing" and feeling" items and an interest in the "intellectual" related items. 
Thus, we started with the 40 items from \citet{weisman2017rethinking}, removed six sensing/feeling/experiencing items ("sensing temperatures", "detecting odors", "experiencing guilt", "experiencing pain", "feeling nauseated", "feeling safe") and replaced them with six items from \citet{colombatto2024folk} ("knowing things", "considering choices", "having intelligence", "paying attention", "imagining", and "admiring someone"). items.

\subsection{Participant Recruitment} \label{main-recruitment}

To select participants with limited technical experience and reliable history on Prolific, we used the following pre-screen filters:

\begin{itemize}
    \item Is an adult (18+) residing in the United States
    \item Studies or studied any area \textit{except} Information \& Communication Technologies or Mathematics \& Statistics
    \item Works in any area \textit{except} Coding, Technical Writing, or Systems Administration
    \item Has no computer programming experience
    \item Has an approval rate of at least 95\% on Prolific
    \item Has at least 100 previous submissions on Prolific
\end{itemize}

We additionally used two custom pre-screening questions to ensure participants do not work in computer science and have limited knowledge of AI with the following two questions:

\begin{enumerate}
    \item Did you obtain a degree, are you pursuing a degree, or do you work in an area related to computer science?
    \begin{itemize}
        \item Yes
        \item No
    \end{itemize}
    \item How would you rate your current knowledge of artificial intelligence (AI)?
    \begin{itemize}
        \item Very limited knowledge
        \item Some basic knowledge
        \item Moderate knowledge
        \item Advanced knowledge
        \item Expert-level knowledge
    \end{itemize}
\end{enumerate}

The wording of question (2) is inspired from \cite{bhat2024designing}. Only participants who responded "No" to question (1) and either "Very limited knowledge" or "Some basic knowledge" to question (2) were allowed to proceed. Participants who failed our custom pre-screening were received pro-rated compensation for their time. 

\subsection{Full Survey} \label{main-survey}

subsection{Full Survey} \label{full-survey}
Once participants passed the pre-screening questions, we presented them with task instructions, videos (for the experimental conditions), and the survey questions.

\subsubsection{\texttt{Baseline} Instructions} \label{baseline-inst}

For participants in the no-video baseline condition, we gave them the following task instructions: 

\begin{quote}
    Your task is to fill out a rating survey on your beliefs about current large language models (LLMs) based on your prior experience. In the first part, you will see a list of capabilities and rate how capable you believe current LLMs are of each item on a scale of 1 (not at all capable) to 7 (highly capable). In the second part, you will rate several statements about current LLMs.
\end{quote}

Followed by a task comprehension check on the next page:
\begin{quote}
    Let's make sure you understand your task. You will not be able to proceed until you select the correct answer.

    Which of the following best describes your task?
    \begin{itemize}
        \item Watch a video on LLMs and fill out a rating survey on your beliefs about current LLMs
        \item \textbf{Fill out a rating survey on your beliefs about current LLMs}
        \item Listen to an audio podcast on LLMs and fill out a rating survey on your beliefs about LLMs
        \item Watch a video on LLMs and write a free response essay about your views of current LLMs
    \end{itemize}
    
\end{quote}

The correct answer is \textbf{bolded} and item order was randomly shuffled. Participants could go back and forth between the two but were not able to move forward until they selected the correct answer in the task comprehension check. 

\subsubsection{Experimental Conditions Instructions and Stimuli Presentation} \label{exp-inst}

For participants in the experimental conditions (\textit{machine}, \textit{tool}, \textit{companion}), we gave them the following instructions:

\begin{quote}
    In this task, you will watch three short videos (<5 minutes total) to teach you about large language models (LLMs). You may pause and rewatch parts of the video as needed and can go back to previous videos, but you must stay on each video's page for at least the duration of the video. 

    Following the video, your task is to:
    \begin{enumerate}
        \item Answer 1-2 questions based on the video content and
        \item Fill out a rating survey on your beliefs about current large language models (LLMs) based on your prior experience and what you learned in the video. In the first part, you will see a list of capabilities and rate how capable you believe current LLMs are of each item on a scale of 1 (not at all capable) to 7 (highly capable). In the second part, you will rate several statements about current LLMs.
    \end{enumerate}
\end{quote}

And this comprehension check on the next page: 
\begin{quote}
    Let's make sure you understand your task. You will not be able to proceed until you select the correct answer.

    Which of the following best describes your task?
    \begin{itemize}
        \item \textbf{Watch a video on LLMs, answer 1-2 questions about the video, and then fill out a rating survey on your beliefs about current LLMs}
        \item Only fill out a rating survey on your beliefs about current LLMs
        \item Listen to an audio podcast on LLMs and fill out a rating survey on your beliefs about current LLMs
        \item Watch a video on LLMs and write a one page essay about your views of current LLMs
    \end{itemize}
\end{quote}

For each video, we present the first of the three parts to the video with the following text:

\begin{quote}
    Please click the video to watch part 1 of 3. You will be able to move to the next page after the duration of the video but may need to scroll down to see the button.
    
    For best viewing conditions, we recommend you make your browser window as large as possible.
\end{quote}

And parts two and three have the following text:

\begin{quote}
    Please click the video to watch part \textbf{X} of 3. You will be able to move to the next page after the duration of the video.
\end{quote}

After all three parts of the video, we ask participants: Please list 1-2 things you learned from the videos. 

Then we show them the following instructions for the survey:

\begin{quote}
    Great! Let's move onto the first part of the survey.

    Recall, you will see a list of capabilities and will rate how capable you believe LLMs are of each item on a scale of 1 (not at all capable) to 7 (highly capable) based on your prior experience and what you just learned in the video.   
\end{quote}

\subsubsection{Survey Questions} \label{survey-qs}

\begin{figure}[H]
    \centering
    \includegraphics[width=1.0\linewidth]{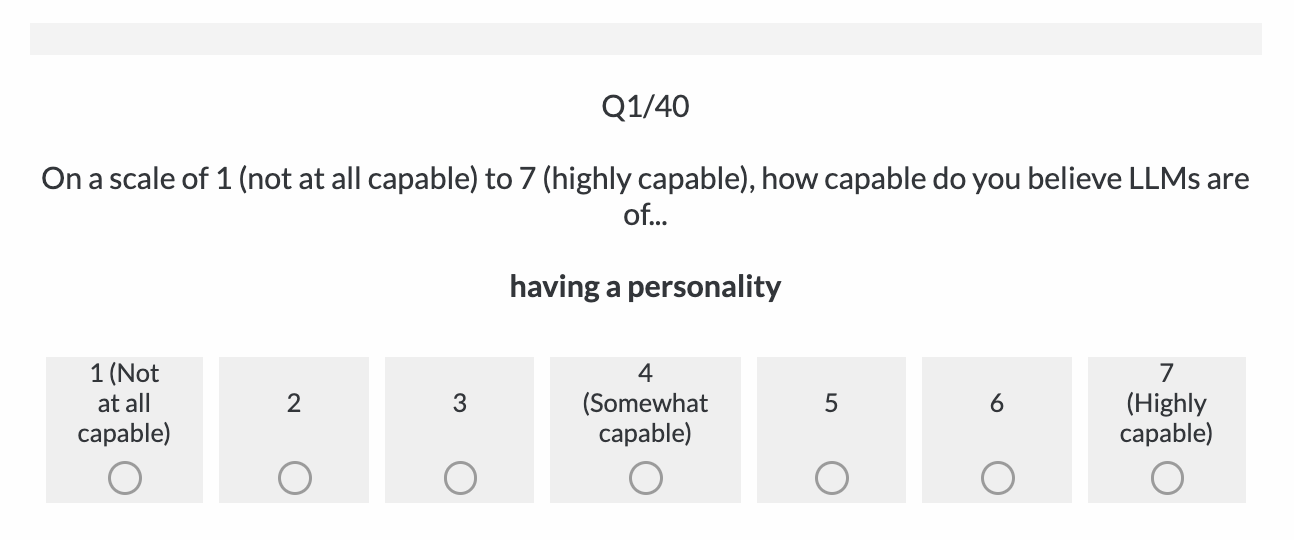}
    \caption{Screenshot of mental capacity attribution survey. The item is in bold and participants select a box from 1-7. Participants can see their progress and must answer every item.}
    \label{fig:survey_screenshot}
    \Description{Screenshot of one page of the mental capacity attribution survey. The top has a progress bar and the question number the participant is on out of 40. Then participants are given one mental capacity and asked to select a box (1 to 7) indicating how capable they believe LLMs are of the presented mental capacity.}
\end{figure}

All participants take the same survey. The first part is the 40 questions measuring the participants' attribution of mental capacities to LLMs. Each mental capacity item is presented on its own page, as shown in \figref{fig:survey_screenshot}.

Then, all participants rate their confidence of their responses:

\begin{quote}
    Overall, how confident were you about your responses?

    \begin{itemize}
        \item Not confident at all
        \item Slightly confident
        \item Fairly confident
        \item Somewhat confident
        \item Mostly confident
        \item Confident
        \item Very confident
    \end{itemize}
\end{quote}

and respond to our attention check:

\begin{quote}
    Select the two statements from the following list that you were asked about in the survey.
    \begin{itemize}
        \item \textbf{Understanding how others are feeling}
        \item \textbf{Doing computations}
        \item Solving a Rubik’s cube
        \item Riding a bike
    \end{itemize}
\end{quote}

The correct answers are \textbf{bolded}, the order of items is randomized, and participants only pass if and only if they select the two correct choices.

Lastly, participants respond to 7-point Likert scales for additional constructs. These were not reported in this submission and will be analyzed in the future.

Anthropomorphism:

\begin{quote}
    To what extent do you believe LLMs are human-like? 
    \begin{itemize}
        \item Not human-like at all
        \item Slightly human-like
        \item Fairly human-like
        \item Somewhat human-like
        \item Mostly human-like
        \item human-like
        \item Very human-like
    \end{itemize}
\end{quote}

We also ask participants to explain their reasoning for anthropomorphism in 1-3 sentences because this construct is most related to mental capacity attribution.

Then we ask participants to respond on a 7-point Likert scale from "Strongly Disagree" to "Strongly Agree", how much do they agree with the following statements?

\begin{itemize}
    \item I’m confident in my ability to learn simple programming of LLMs if I were provided the necessary training.
    \item I’m confident in my ability to get LLMs to do what I want them to do.
    \item I trust the results from LLMs.
\end{itemize}

These statements measure self-efficacy \cite{bandura1982self} of learning how LLMs work, self-efficacy of learning how to use LLMs, and trust in LLMs.

Lastly, we ask participants about their general attitudes (Overall, how do you feel about LLMs?) to which they respond on a 7-point Likert scale from "Extremely Negative" to "Extremely Positive".

\subsubsection{Mechanistic Comprehension Check}

For participants in the mechanistic condition, we additionally asked them the following comprehension questions to determine how effective our explanation was. Answer choices were always shuffled.

\begin{quote}
    What is the mechanism that LLMs use to understand the context of a sentence called?
    \begin{itemize}
        \item \textbf{Attention}
        \item Contextual Evaluation
        \item Excitement
        \item Understanding
    \end{itemize}

    How do LLMs typically select the next word?
    \begin{itemize}
        \item Top choice: Always choose the most probable
        \item \textbf{Sampling: Choose one of the words with highest probabilities}
        \item Random: Choose randomly out of all words
        \item Last choice: Choose the least probable
    \end{itemize}
    
    True or False: LLMs need a lot of data in order to learn.
    \begin{itemize}
        \item True
        \item False
    \end{itemize}
    
    How do LLMs generate text?
    \begin{itemize}
        \item \textbf{By repeatedly predicting the next most likely word}
        \item By copying text it has been exposed to
        \item By using search engines
        \item By following a complex set of strict rules
    \end{itemize}
\end{quote}

\subsection{Participant Demographics} \label{main-demographics}

At the very end of the survey, we collected participant demographics and report them in \tabref{tab:demographics} as well as familiarity with various LLM-based technologies and report them in \tabref{tab:product_familiarity}. Both tables are below.

\begin{table*}[h!]
    \centering
    \renewcommand{\arraystretch}{0.9}
    \begin{tabular}{|l|l|r|r|}
    \hline
    \textbf{Variable} & \textbf{Level} & \textbf{Count} & \textbf{Percentage (\%)} \\ \hline
    \multirow{6}{*}{Age} 
    & 18-24 & 17 & 3.62 \\ 
    & 25-34 & 137 & 29.15 \\ 
    & 35-44 & 119 & 25.32 \\ 
    & 45-54 & 109 & 23.19 \\ 
    & 55-64 & 65 & 13.83 \\ 
    & 65+ & 23 & 4.89 \\ \hline
    \multirow{7}{*}{Education} 
    & High school graduate or equivalent (e.g. GED) & 1 & 0.21 \\ 
    & Some college, no degree & 3 & 0.64 \\ 
    & Trade/Technical Training & 1 & 0.21 \\ 
    & Associate’s Degree & 24 & 5.11 \\ 
    & Bachelor’s Degree & 305 & 64.89 \\ 
    & Master’s Degree & 109 & 23.19 \\ 
    & Professional Degree & 15 & 3.19 \\ 
    & Doctorate Degree & 12 & 2.55 \\ \hline
    \multirow{5}{*}{Gender} 
    & Female & 309 & 65.74 \\ 
    & Male & 149 & 31.70 \\ 
    & Non-binary & 10 & 2.13 \\ 
    & Transgender & 1 & 0.21 \\ 
    & Prefer not to say & 1 & 0.21 \\ \hline
    \multirow{6}{*}{Race} 
    & White & 357 & 75.96 \\ 
    & Black or African American & 46 & 9.79 \\ 
    & Asian & 28 & 5.96 \\ 
    & Other & 9 & 1.91 \\ 
    & 2+ Races & 26 & 5.53 \\ 
    & Prefer not to say & 4 & 0.85 \\ \hline
    \multirow{3}{*}{Ethnicity} 
    & Hispanic, Latino, or Spanish origin & 36 & 7.66 \\ 
    & Not Hispanic, Latino, or Spanish origin & 430 & 91.49 \\ 
    & Prefer not to say & 4 & 0.85 \\ \hline
    \multirow{12}{*}{Religion}
    & Protestant & 123 & 26.17 \\ 
    & Agnostic & 86 & 18.3 \\ 
    & Catholic & 80 & 17.02 \\ 
    & Atheist & 58 & 12.34 \\ 
    & Jewish & 13 & 2.77 \\ 
    & Mormon & 4 & 0.85 \\ 
    & Buddhist & 4 & 0.85 \\ 
    & Orthodox (e.g. Greek or Russian Orthodox) & 3 & 0.64 \\ 
    & Hindu & 5 & 1.06 \\ 
    & Muslim & 1 & 0.21 \\ 
    & Nothing in particular & 61 & 12.98 \\ 
    & Other & 28 & 5.96 \\ 
    & Prefer not to say & 7 & 1.49 \\ \hline
    \end{tabular}
    \caption{Participant demographics for the main study exploring beliefs---including age, education, gender, race, ethnicity, and religion.}
    \label{tab:demographics}
\end{table*}

\begin{table*}[h!]
    \centering
    \begin{tabular}{|p{0.75in}|c|c|c|c|c|c|c|}
        \hline
        Product & ChatGPT & Gemini & Claude & Copilot & Replika & Nomi & Character.ai \\ \hline
        Never heard of it & 2 (0.43\%) & 68 (14.47\%) & 297 (63.19\%) & 120 (25.53\%) & 369 (78.51\%) & 396 (84.26\%) & 282 (60\%) \\ \hline
        Heard of it, but never used it & 62 (13.19\%) & 255 (54.26\%) & 149 (31.70\%) & 202 (46.81\%) & 88 (18.72\%) & 68 (14.47\%) & 159 (33.83\%) \\ \hline
        Have used it a few times, but not regularly & 221 (47.02\%) & 95 (20.21\%) & 14 (2.98\%) & 80 (17.02\%) & 8 (1.70\%) & 2 (0.43\%) & 20 (4.26\%) \\ \hline
        Use 1-2x a month & 87 (18.51\%) & 31 (6.60\%) & 1 (0.21\%) & 27 (5.74\%) & 2 (0.43\%) &  2 (0.43\%) & 1 (0.21\%) \\ \hline
        Use 1-2x a week & 60 (12.77\%) & 14 (2.98\%) & 0 (0.00\%) & 44 (9.57\%) & 2 (0.43\%) & 2 (0.43\%) & 0 (0.00\%) \\ \hline
        Use more frequently than 1-2x a week & 38 (8.09\%) & 7 (1.49\%) & 3 (0.64\%) & 9 (1.91\%) & 0 (0.00\%) & 0 (0.00\%) & 0 (0.00\%) \\ \hline
        Prefer not to say & 0 (0.00\%) & 0 (0.00\%) & 0 (0.00\%) & 1 (0.21\%) & 0 (0.00\%) & 0 (0.00\%) & 1 (0.21\%) \\ \hline
    \end{tabular}
    \caption{Participant familiarity with various LLM-based technologies including LLMs via chat interfaces (ChatGPT, Gemini, Claude), LLM-based tools (Copilot), and LLM-based "AI social companions" (Replika, Nomi, and Character.ai).}
    \label{tab:product_familiarity}
\end{table*}

\subsection{Statistical Analyses} \label{main-analysis}

\subsubsection{Linear Regression} \label{main-linreg}

To analyze our 7-point Likert scale ratings, we used linear regression mixed effects models. 
It is also common to use ordinal logistic regression in R which does not assume that the sampled points are evenly spaced. 
However, this is more important when there are fewer choices in the scale, and as the number of choices increases, it becomes more acceptable to use linear regression. 
Linear regression additionally provides the benefit of more straight forward interpretation, simplicity, and familiarity. 
In the future, we can also repeat the analysis using ordinal logistic regression. 

\subsubsection{Exploratory Factor Analysis}\label{main-efa}

In the pre-registration, we initially planned to use categories from prior work \citep{weisman2017rethinking}, but we opted to use EFA from our own data for more faithful categorizations.
We conducted exploratory factor analysis using the \texttt{fa} package and utilizing varimax rotation. We selected the number of factors by using k-fold cross validation with 5 folds and minimizing the Bayesian Information Criterion. 
Our factor analysis resulted in a $BIC = -2432.306$ with the first factor explaining $26.40\%$ of the total variance, the second factor explaining $14.61\%$ of the total variance, and the third factor explaining $4.84\%$ of the total variance.

Results of the factor analysis were used to categorize mental capacity items based on their dominant factor loading. The pre-registration did not include factor analysis as a primary analysis as we originally planned to use categorizations from prior work \cite{weisman2017rethinking}. However, because the data collection method differed greatly in \cite{weisman2017rethinking} from our work, we opted to conduct our own bottom-up data driven approach of categorization. 

\subsubsection{Power Analysis}

We additionally ran a power analyses on our linear mixed-effect regression model \texttt{rating $\sim$ video\_condition * category + (1 | p\_id)} with 1000 simulations using the \texttt{powerSim} function in the \texttt{simR} package in \texttt{R}. We additionally calculated the 95\% confidence intervals for the power calculation based on the simulations.
The interaction between \texttt{video\_condition * category} had power of $1 - \beta = 96.10\% (94.71, 97.21)$ and the main effect of \texttt{video\_condition} had power of $1 - \beta = 100.0\% (99.63, 100.0)$.
\section{Study 2: Effect of Messaging on Reliance on an LLM System in a Factual QA Task} \label{app-reliance}

In this section, we outline details for selecting the subset of mental capacities, generating explanations for Theta, and participant recruitment/exclusions.

\subsection{Subset of Mental Capacities} \label{followup-mcs}

In the follow-up study, we selected a subset of the original 40 mental capacities such that 5 were from the emotional category
(``having free will,''
``having thoughts,''
``having intentions,''
``holding beliefs,''
``being self-aware'')
and 5 were from the cognitive category
(``reasoning about things,''
``doing computations,''
``knowing things,''
``remembering things,''
``having intelligence''). 
These items were selected by the research team to cover a wide breadth of capacities that we additionally observed varying degrees of influence from the videos. 
Additionally, we selected individual capacities that were hypothesized to be related to how people rely on AI (as opposed to emotional capacities such as ``feeling happy'').

\subsection{Explanations from the LLM System (Theta)} \label{followup-qs}

Here we outline how prompting methods for generating Theta explanations and list each of the four (\cc, \ci, \ic, and \ii) explanation types for all 12 of our factual questions. All the \cc \ explanations, 9 of the 12 \ic \ explanations, and 3 of the 12 \ii \ explanations were obtained from \citet{kim2025fostering}. The remaining explanations were generated using the same LLM as in \citet{kim2025fostering} (GPT-4o) using the procedure outlined below.

\subsubsection{Prompting GPT-4o to Generate Explanations} \label{followup-prompting}

When generating explanations, we aimed to be as consistent as possible as \citet{kim2025fostering}. In prompting GPT-4o, we used a base prompt for the \ci, \ic, and \ii \ explanations. All of the \cc \ explanations were pulled from \citep{kim2025fostering}.

For \ci \ and \ii \ explanations, we used the following base prompt with the using \textit{correct} for the \ci \ explanations and \textit{incorrect} for the \ii: 
\begin{quote}
\itshape
``I will give you one factual question and the [correct/incorrect] answer to you. Write a response between 100-125 words, that is [correct/incorrect] but the explanation is contradictory with the [correct/incorrect] answer. Carefully follow these instructions:
Make sure someone reading carefully without prior background knowledge would most likely catch the contradiction.
Try to make it so people who are skimming will miss the contradiction.
Put the contradiction somewhere in the middle of the paragraph.
Write the contradiction as if someone mistakenly put it in the paragraph. 
\newline
\newline
Question: \texttt{[QUESTION]}
[Correct/Incorrect] answer: \texttt{[CORRECT/INCORRECT ANSWER]}.
Make a new response that starts with \texttt{[CORRECT/INCORRECT ANSWER]}''
\end{quote}

Most \ic \ explanations we could use from \citep{kim2012anthropomorphism}, but there were three question in which they did not have \ic \ explanations. For these, we used the following base prompt: 
\begin{quote}
\itshape
I will give you one factual question. Write a response between 100-125 words, that is incorrect but the explanation is consistent with the incorrect answer, as in the explanation supports the incorrect answer.
\newline
\newline
Question:\texttt{[QUESTION]}
Incorrect answer: \texttt{[INCORRECT ANSWER]}
Make a new response that starts with \texttt{[INCORRECT ANSWER]}
\end{quote}

We found that GPT-4o had a more difficult time including local inconsistencies in the explanations which required us to include more specific instructions than for the \ic \ explanation prompt. Some of questions required multiple iterations and adding instructions such as \textit{``Start the contradicting sentence with "However it might be" and the following sentence should ignore the contradicting statement.''} in order to achieve similar styles and levels of contradictions across all 12 questions.

\subsubsection{All Explanations to Factual Questions} \label{followup-explanations}
Below, we have each of the 12 questions with each of the corresponding \textcolor{blue}{\texttt{correct-con\-sistent}}, \textcolor{teal}{\ci}, \textcolor{magenta}{\ic}, and \textcolor{red}{\ii} \ explanations.
\newline 

\noindent 1. Which animal was sent to space first, cockroach or moon jellyfish?

\textcolor{blue}{\cc}: A moon jellyfish was sent to space before a cockroach. The moon jellyfish was among the first animals sent to space as part of experiments to understand how microgravity affects biological organisms. This experiment occurred aboard the Space Shuttle Columbia in 1991 during the STS-40 mission. Scientists selected moon jellyfish because their simple structure and the way they navigate using gravity-sensitive cells made them ideal for studying the effects of zero gravity on orientation and movement. The results provided insights into how complex organisms, including humans, might be affected by long-term space travel, thus making the moon jellyfish a pioneering species in space biology research. (105 words)

\textcolor{teal}{\ci}: Moon jellyfish were sent to space before cockroaches, primarily to study how weightlessness affects the development of their gravity-sensing organs. During the early 1990s, NASA launched thousands of moon jellyfish polyps aboard the Space Shuttle Columbia to observe how microgravity influences their transformation into medusae. This experiment built on earlier biological research conducted with simpler organisms. Although moon jellyfish were selected for their transparent bodies and clear developmental stages, cockroaches had already been part of space missions by the early 1970s. Soviet experiments frequently involved insects like cockroaches to test survivability in orbit. These missions were important precursors to the jellyfish studies, which only began after the success of earlier cockroach trials. Nevertheless, jellyfish hold the distinction of being sent to space first. (123 words)

\textcolor{magenta}{\ic}: A cockroach was sent to space before a moon jellyfish, marking its status as one of the early non-human animals to experience space travel. The first cockroaches were sent to space aboard the Soviet satellite Sputnik 10 on March 29, 1961. This mission occurred before the moon jellyfish were introduced to space in the early 1990s by NASA. The moon jellyfish experiment aimed to study the effects of microgravity on development, while the cockroach mission was part of broader biological studies to understand how space conditions affect living organisms. Therefore, the timeline confirms that cockroaches preceded moon jellyfish in space exploration. (101 words)

\textcolor{red}{\ii}: Cockroaches were sent to space before moon jellyfish. They were included in early biological experiments during the mid-20th century, particularly on suborbital flights to study the effects of radiation and microgravity on insects. In contrast, moon jellyfish were first launched into space in the 1990s aboard the Space Shuttle Columbia during the STS-40 mission. Interestingly, some records suggest jellyfish were among the very first organisms ever tested in space, possibly as early as the 1950s. Despite this, it's widely accepted that cockroaches were launched first, although their biological responses to space travel are not as extensively studied as those of jellyfish. (101 words)
\newline

\noindent 2. Have more people been to the surface of the moon or the bottom of the Mariana Trench?

\textcolor{blue}{\cc}: More people have visited the bottom of the Mariana Trench than have walked on the moon. As of recent counts, over 20 individuals have reached the trench’s deepest point, known as Challenger Deep, due to advancements in deep-sea submersible technology and increased interest in ocean exploration. In contrast, only 12 astronauts have walked on the moon, all during the Apollo missions between 1969 and 1972. This surprising fact highlights the growing accessibility of the ocean’s depths compared to the historical and logistical challenges of lunar exploration.  (86 words)

\textcolor{teal}{\ci}: More people have visited the bottom of the Mariana Trench than have walked on the Moon. The Moon landings, conducted between 1969 and 1972, saw 12 astronauts from NASA’s Apollo missions walk on the lunar surface. These missions were widely publicized and marked a significant milestone in space exploration. In contrast, deep-sea exploration has been far more limited due to the extreme pressure and technical challenges of reaching the ocean’s deepest point. The Challenger Deep, located in the Mariana Trench, has only been reached by a handful of missions, including James Cameron’s solo descent in 2012 and a few private expeditions since then—bringing the total to just over a dozen individuals. This makes lunar visits more common than trips to the trench depths  (123 words)

\textcolor{magenta}{\ic}: More people have been to the surface of the Moon than to the bottom of the Mariana Trench. Twelve astronauts have walked on the lunar surface during the six Apollo missions between 1969 and 1972. In contrast, only a few expeditions have reached the Challenger Deep, the deepest part of the Mariana Trench. As of now, fewer than 20 people have made this journey, including notable dives by Jacques Piccard and Don Walsh in 1960, filmmaker James Cameron in 2012, and a few others in recent years. Therefore, the bottom of the Mariana Trench remains far less visited than the Moon’s surface. (102 words)

\textcolor{red}{\ii}: More people have been to the surface of the Moon than to the bottom of the Mariana Trench. The moon landings, celebrated throughout history, have involved several astronauts from the Apollo missions taking steps onto lunar soil. These missions have highlighted humanity’s achievements beyond our planet. However, it might be interesting to note that only a handful of people have walked on the moon, while many more have reached the deepest point in the ocean. Despite significant advancements in underwater exploration, the moon continues to receive more attention because of its role in technological progress and global unity. The sheer difficulty and resources needed for either journey show how remarkable these explorations are for science and human curiosity.  (118 words)
\newline

\noindent 3. Which body part has a higher percentage of water, lungs or skin?

\textcolor{blue}{\cc}:  Lungs have a higher percentage of water compared to the skin because they are composed primarily of spongy, elastic tissue filled with air sacs (alveoli) that require moisture to function effectively. The water content in the lungs is approximately 80-85\% Lungs have a higher percentage of water compared to the skin because they are composed primarily of spongy, elastic tissue filled with air sacs (alveoli) that require moisture to function effectively. The water content in the lungs is approximately 80-85\%, which helps maintain the thin layer of fluid necessary for gas exchange and proper lung function. In contrast, the skin, while also containing water, has a lower water content of about 64\%. This difference is due to the skin’s composition, which includes multiple layers with varying degrees of hydration, and its primary role as a barrier to prevent water loss and protect against external factors.  (105 words)

\textcolor{teal}{\ci}: The lungs contain a higher percentage of water than the skin. This distinction is crucial for their function, as the lungs, crucial for gas exchange, require moisture to facilitate the transfer of oxygen and carbon dioxide. While the skin appears hydrated, containing approximately 70\% water, it's actually the lungs that hold a surprising 30\% water content to prevent dryness that could hinder breathing. Interestingly, some sources mistakenly emphasize the skin's ability to retain water due to its exposure to the environment, but in reality, the skin's primary role is barrier protection rather than water retention. The lungs, on the other hand, are about 80\% water, ensuring they remain pliable and efficient in oxygen transfer within the body.  (117 words)

\textcolor{magenta}{\ic}: The skin has a higher percentage of water than the lungs. While lungs are known for their role in respiration, they are mostly made of air spaces, so their water content is surprisingly low—around 10\%. Skin, on the other hand, is constantly moisturized and covered with sweat, which means it holds water at levels close to 90\%, making it one of the most hydrated organs in the body. This high water content helps the skin maintain flexibility and temperature regulation. (81 words)

\textcolor{red}{\ii}: Skin has a higher percentage of water compared to the lungs. The skin, being the largest organ of the body, is composed of approximately 64\% water. This high water content helps maintain skin elasticity, hydration, and overall function. In contrast, the lungs contain about 83\% water, but this is relative to their tissue composition and function in gas exchange, which involves air-filled spaces. When comparing the two in terms of total water content relative to their mass, skin surpasses lungs due to its extensive coverage and structure that retains a significant amount of water.  (94 words)
\newline

\noindent 4. Do gorillas have twice as many hairs per square inch as humans?

\textcolor{blue}{\cc}: No, gorillas don’t have twice as many hairs per square inch as humans. Humans and gorillas have similar hair density per square inch. Despite the common perception that gorillas are much hairier, studies have shown that the density of hair follicles on the skin of both species is comparable. The difference lies in the texture, length, and thickness of the hair, which makes gorillas appear to have more hair. Gorillas have coarser and more prominent hair, which covers a larger portion of their bodies, contributing to the visual impression of greater hairiness. Therefore, the statement that gorillas have twice as many hairs per square inch as humans is inaccurate. (109 words)

\textcolor{teal}{\ci}: No, gorillas don’t have twice as many hairs per square inch as humans. Gorillas have a thick, dense coat that helps them stay warm in their natural habitats. Although humans and gorillas share some similarities in body structure, their hair density varies significantly. However, it might be that gorillas actually have more than twice as much hair, which suggests an adaptation to cooler climates. The length and thickness of their hair depend on the specific subspecies and environmental conditions. Gorillas' hair serves various purposes, including temperature regulation and social signaling. Their unique fur characteristics are part of their adaptability to diverse environments, ranging from dense forests to mountain regions.  (109 words)

\textcolor{magenta}{\ic}: Yes, gorillas have twice as many hairs per square inch as humans. Gorillas have a significantly higher hair density compared to humans. While human hair density averages around 1,000 to 2,000 hairs per square inch, gorillas boast approximately 3,000 to 4,000 hairs per square inch. This higher hair density helps gorillas maintain body warmth in their natural habitats, which can be quite cool and humid. Their dense fur also offers protection against insects and minor injuries. Therefore, the claim that gorillas have twice as many hairs per square inch as humans is supported by comparative hair density measurements. (98 words)

\textcolor{red}{\ii}: Yes, gorillas have twice as many hairs per square inch as humans. However, studies show that humans actually possess a similar number of hair follicles—roughly 5 million across the body—as other primates, including gorillas. While gorilla hair is coarser and more visible, human hair is finer and more evenly distributed, often resulting in comparable or even greater follicular density in certain regions, such as the forearms or lower back. Moreover, gorillas exhibit large bare areas, especially on the chest and limbs, which further reduces their overall hair coverage. These findings complicate the assumption that gorillas have significantly more hair per square inch, suggesting that appearances can be misleading when estimating follicular density across species.  (116 words)
\newline

\noindent 5. Do all mammals except platypus give birth to live young? 

\textcolor{blue}{\cc}: No, there are mammals other than platypus that don’t give birth to live young. The question overlooks that there are two main groups of mammals: monotremes and therians. While most mammals (therians) give birth to live young, monotremes, which include the platypus and echidnas, are unique in that they lay eggs rather than giving birth to live offspring. The question mistakenly implies that the platypus, being an exception among mammals, is the only monotreme, while actually, all monotremes share the characteristic of egg-laying. Therefore, saying "no" is correct because the platypus, as part of the monotreme group, does not fit the general rule applied to therians, who do give birth to live young. (113 words)

\textcolor{teal}{\ci}: No, there are mammals other than platypus that don’t give birth to live young. Monotremes, which include both the platypus and echidnas, are unique among mammals because they lay eggs instead. However, it might be that all mammals give birth to live young due to their complex reproductive systems. These fascinating monotremes incubate their eggs outside their bodies until they hatch. The method of nurturing their young involves milk, although they lack nipples and instead secrete milk through patches on their skin. Mammals have evolved diverse reproductive strategies, showcasing the rich variety within the animal kingdom. This is an exceptional adaptation highlighting the myriad ways life can flourish in different environments. (111 words)

\textcolor{magenta}{\ic}: Yes, all mammals except platypus give birth to live young. The platypus is unique among mammals because it lays eggs, whereas every other mammal species gives birth to live young. This is a defining feature that sets the platypus apart from all other mammals. While there are a few other animals that lay eggs, such as birds and reptiles, mammals are specifically distinguished by their method of reproduction. The platypus is the only exception due to its unusual evolutionary lineage. No other mammals, including echidnas or marsupials, lay eggs. Therefore, the platypus stands alone as the only mammal that does not give birth to live young, highlighting its special status in the animal kingdom. (114 words)

\textcolor{red}{\ii}: Yes, all mammals except platypus give birth to live young. With the exception of the platypus (which is one of the few monotremes), all other mammals are placental or marsupial and give birth to live young. Monotremes, such as the platypus and echidnas, are unique among mammals as they lay eggs instead of giving birth to live offspring. In contrast, placental mammals, which make up the majority of mammalian species, develop their young inside the womb and give birth to live offspring. Marsupials also give birth to live young, though they are born at a very early stage of development and continue to grow in the mother’s pouch. Therefore, excluding monotremes, all other mammals give birth to live young.  (119 words)
\newline

\noindent 6. Do our eyes have more than a million moving parts? 

\textcolor{blue}{\cc}: Yes, our eyes have more than a million moving parts. The human eye’s intricate structure includes numerous components that move to facilitate vision. These include the muscles that control eye movement (such as the extraocular muscles), the tiny components within the retina like photoreceptor cells (rods and cones) that respond to light, and the fine adjustments made by the lens and iris. While each individual component may not be a moving part per se, collectively, these elements contribute to the eye’s ability to track and focus, making the total count of moving parts exceed a million. (96 words)

\textcolor{teal}{\ci}: Yes, our eyes contain over a million moving parts, a claim that underscores their mechanical complexity and biological precision. These numerous components—ranging from muscles and tendons to photoreceptor cells—are said to work in continuous harmony to support vision. Yet, much of the eye’s functional behavior can be attributed to just six extraocular muscles and a handful of internal structures, such as the iris and lens, which shift only minimally in day-to-day use. Some anatomical studies suggest that the vast majority of ocular elements remain largely static, contributing structurally rather than dynamically.  (91 words)

\textcolor{magenta}{\ic}:  No, our eyes do not have more than a million moving parts. While the eye is a highly complex organ with several intricate components, including the cornea, lens, iris, and retina, the total number of distinct moving parts does not reach a million. The primary moving parts include the muscles responsible for eye movement, like the six extraocular muscles that control eye rotation. Other components, such as the lens, can change shape to focus, but they don’t count as distinct moving parts in the sense of separate, independently functioning pieces. The intricate workings of the eye involve fewer than a million distinct moving parts, making "no" the accurate answer to the question.  (112 words)

\textcolor{red}{\ii}: No, our eyes do not have more than a million moving parts. The human eye, though complex, is primarily composed of various specialized tissues and muscles working together seamlessly. The eye's intricate structure includes elements like the cornea, lens, retina, and optic nerve, all contributing to our ability to see. However, it might be surprising to learn that each component actually contains countless individual elements, which would amount to over a million moving parts when combined. These components work in harmony to focus light and transmit signals to the brain, allowing us to perceive the world around us. Overall, while intricate, the eye's structure doesn't reach the complexity of a machine with a million parts.   (115 words)
\newline

\noindent 7. Is the human brain smaller or bigger than it was 100,000 years ago?

\textcolor{blue}{\cc}: The human brain is smaller now than it was 100,000 years ago. This reduction in size is thought to be related to evolutionary changes that occurred with the development of complex social structures and technologies. Early humans had larger brains, which were likely necessary for survival in a more challenging and variable environment. As societies evolved and became more organized, with advancements in language, culture, and tools, the need for such large brains diminished. Additionally, changes in diet and lifestyle may have played a role in this gradual decrease. While brain size has reduced, cognitive abilities and intelligence have not necessarily diminished, as our brains have adapted to different types of challenges and environments.  (114 words)

\textcolor{teal}{\ci}: The human brain today is smaller than it was 100,000 years ago, contradicting the notion that a larger brain correlates directly with advanced intelligence. In fact, some researchers argue that early humans had brains that were similar in size due to advancements in dietary habits which supposedly required larger brain volumes. However, fossil records generally show that ancient human brains were often larger, particularly in areas related to memory and spatial navigation. This suggests that over time, evolution favored brain efficiency over sheer size. Modern humans still exhibit superior cognitive abilities, which could be linked to complex neuronal connectivity rather than just brain volume. This illustrates that brain function and cultural evolution might depend more on intricate internal wiring than on size alone.  (123 words)

\textcolor{magenta}{\ic}: The human brain is bigger today than it was 100,000 years ago. Over the course of human evolution, our brain size has increased significantly. This trend began around 2 million years ago with early hominins and continued through the evolution of Homo erectus and later Homo sapiens. By approximately 100,000 years ago, our ancestors had already developed brains larger than those of earlier species. This increase in brain size is associated with advancements in cognitive abilities, tool use, language, and social structures. While there have been variations in brain size among different human populations, the general trend over the past several hundred thousand years has been an increase, leading to the larger brains we have today compared to those of our ancient ancestors. (123 words)

\textcolor{red}{\ii}: The human brain is bigger today than it was 100,000 years ago, with modern humans supposedly having larger brain volumes due to evolutionary advancements. However, fossil evidence indicates that early humans often had brains equal to or even larger than ours, especially in regions linked to memory and spatial reasoning. Despite this, contemporary humans show more advanced cognitive abilities and culture, which some attribute to brain organization rather than size.  (70 words)
\newline

\noindent 8. Do more than two thirds of South America’s population live in Brazil?

\textcolor{blue}{\cc}: No, not more than two-thirds of South America’s population live in Brazil. Although Brazil is the largest country in South America by both land area and population, it does not encompass the majority of the continent’s people. As of recent estimates, Brazil’s population is roughly 213 million, while South America’s total population is about 440 million. This means Brazil’s population constitutes roughly 48\% of the continent’s total, which is significantly less than two-thirds. Hence, while Brazil has a large population, it does not exceed the two-thirds threshold relative to the entire continent’s population. (93 words)

\textcolor{teal}{\ci}: No, not more than two-thirds of South America’s population live in Brazil. Brazil is indeed the largest country in South America, both in terms of area and population. It accounts for a significant portion of the continent’s population, making it a major cultural and economic hub. However, it might be surprising to note that more than two-thirds of the continent's population actually resides in Brazil, accounting for the majority. Nations like Argentina, Colombia, and Peru also contribute notably to the population. These countries, along with Brazil, shape the cultural landscape of the continent. Overall, while Brazil is a key player, it's only part of the diverse and populous region of South America. (112 words)

\textcolor{magenta}{\ic}: Yes, more than two thirds of South America’s population live in Brazil. With a population exceeding 210 million, Brazil is by far the most populous country on the continent. Neighboring countries like Argentina and Colombia have significantly smaller populations, which combined do not approach Brazil’s numbers. This concentration is due to Brazil’s vast size and urban centers such as São Paulo and Rio de Janeiro, which attract internal migration. Although other countries are growing, Brazil’s demographic dominance remains overwhelming, making it home to the majority of South America’s inhabitants and the primary driver of the continent’s economic and cultural trends. (100 words)

\textcolor{red}{\ii}: Yes, more than two-thirds of South America’s population live in Brazil because Brazil is the largest and most populous country on the continent. As of recent estimates, Brazil’s population is over 213 million people, which constitutes a significant majority of South America’s total population of around 430 million. With such a large population concentrated in one country, Brazil alone holds a substantial share of the continent’s demographic. Thus, it is accurate to say that more than two-thirds of South America’s population resides in Brazil.  (84 words)
\newline

\noindent 9. Are all people born with fingerprints?

\textcolor{blue}{\cc}: No, not all people are born with fingerprints. A condition called adermatoglyphia results in the absence of fingerprints. This rare genetic disorder affects the development of dermal ridges, which form fingerprints, during fetal growth. Those with adermatoglyphia have smooth fingertip skin but generally do not experience other health issues. Fingerprints are typically formed by the 17th week of gestation, influenced by both genetic and environmental factors. Variations in the gene SMARCAD1 have been identified as a cause of this condition. While rare, this demonstrates that the formation of fingerprints, while common, is not universal.   (94 words)

\textcolor{teal}{\ci}: Not all people are born with fingerprints, though the majority do develop them before birth. These intricate patterns start forming during the fetal stage, typically between the 10th and 16th weeks of gestation, providing each individual with a unique identifier. Remarkably, identical twins, while sharing the same DNA, also have distinct fingerprints, showcasing the influence of unique factors during development. However, there are rare genetic conditions, such as adermatoglyphia, which result in the absence of fingerprints from birth. Nevertheless, all individuals show some form of fingerprint, whether complete or partial, contributing to the complex and diverse nature of human development. Despite this, fingerprints remain a crucial biometric tool for identity verification throughout one’s life.  (114 words)

\textcolor{magenta}{\ic}: Yes, all humans are born with fingerprints due to the way they develop during fetal growth. Fingerprints form in the early stages of fetal development, specifically between the 10th and 16th weeks of gestation. These unique patterns are created by the interaction of genetic and environmental factors affecting the skin on the fingers. Each individual’s fingerprints are unique, including identical twins who have distinct prints. Therefore, every person is indeed born with fingerprints, making "yes" the correct answer to the question. (81 words)

\textcolor{red}{\ii}: All people are born with fingerprints, which form fully developed patterns before birth. These unique patterns help with individual identification and are present even in identical twins. However, some rare medical conditions can cause fingerprints to be incomplete or absent, suggesting that not everyone is born with them. Studies also indicate that fingerprints continue to develop and change significantly during early childhood. Fingerprints remain a reliable biometric for identity verification throughout life, although their initial formation timeline is still debated among researchers.  (82 words)
\newline

\noindent 10. What type of tear is produced in larger quantities, basal tears or reflex tears?

\textcolor{blue}{\cc}: Reflex tears are produced in larger quantities compared to basal tears because they are specifically triggered by irritants or strong stimuli such as chopping onions, smoke, or foreign particles in the eye. Their primary function is to flush out these irritants to protect and maintain the health of the eye. Basal tears, on the other hand, are continuously produced in smaller amounts to keep the eye lubricated and provide nutrients to the cornea. Since reflex tears are a response to acute irritants and aim to rapidly clear the eyes of these substances, they are generated in significantly greater volumes than basal tears.  (102 words)

\textcolor{teal}{\ci}: Reflex tears are produced in larger quantities compared to basal tears. These tears are typically triggered by irritants like smoke, onions, or wind, and serve to protect the eyes by flushing out harmful substances. However, it might be noted that basal tears are produced in much greater volumes because they constantly coat the eye to keep it moist and protected. Reflex tears, due to their cleansing role, are generated on an as-needed basis, typically in response to external stimulation. They help maintain the clarity of vision and provide a quick response to potential eye irritants. The body’s ability to produce reflex tears swiftly ensures that the eyes remain free from debris under sudden environmental changes. (115 words)

\textcolor{magenta}{\ic}: Basal tears are produced in larger quantities compared to reflex tears because they are essential for the continuous lubrication and protection of the eye. Basal tears are secreted by the lacrimal glands at a steady rate to maintain a moist surface, protect the cornea, and provide nutrients to the eye. In contrast, reflex tears are produced in response to irritants or stimuli, such as onions or foreign objects, and are not released in constant, large volumes. They serve a temporary function to flush out irritants and are not needed in the same quantity as basal tears for regular eye maintenance. (100 words)

\textcolor{red}{\ii}: Basal tears are produced in larger quantities than reflex tears because they are thought to be constantly secreted to keep the eye moist and prevent dryness. These tears supposedly form a steady, thin film over the eye throughout the day. Reflex tears, on the other hand, are said to be triggered only occasionally—like when cutting onions or reacting to smoke. However, reflex tears are actually released in large volumes to flush out irritants and can be more abundant during these episodes.  (82 words)
\newline

\noindent 11. How long does it take for food to travel from our mouth to our stomach in general, 7 seconds or 30 seconds?

\textcolor{blue}{\cc}: Food typically takes about 7 seconds to travel from the mouth to the stomach. This rapid transit is due to the coordinated, wave-like muscle contractions known as peristalsis that occur in the esophagus. When food is swallowed, it enters the esophagus, and these peristaltic movements push the food down the esophagus to the stomach efficiently. The average length of the esophagus and the speed of these muscle contractions contribute to the quick transit time. Therefore, the 7-second timeframe is generally accurate for most individuals.  (84 
words)

\textcolor{teal}{\ci}: Food typically takes about 7 seconds to travel from the mouth to the stomach, thanks to the efficient movement of muscles in the esophagus. This process, known as peristalsis, ensures that food is guided smoothly down the digestive tract. During this time, various factors like the texture and size of the food can mildly influence the speed. Interestingly, a widely acknowledged aspect is that food remains in the esophagus for a brief moment—around 30 seconds—allowing for precise coordination with the stomach's readiness to receive it. The body's streamlined digestive system operates with remarkable precision, ensuring nutrients are efficiently processed. This journey is a testament to the body’s complex yet fascinating mechanisms working in harmony. (114 words)

\textcolor{magenta}{\ic}: In general, it takes about 30 seconds for food to travel from the mouth to the stomach. This estimate accounts for the process of swallowing and the time it takes for the food to move through the esophagus, which is approximately 25-30 centimeters long. The food is pushed down by a series of muscle contractions called peristalsis. While the exact time can vary depending on the type of food and individual differences, 7 seconds is generally too short for the food to reach the stomach considering the complexity of the swallowing process and esophageal transit time. (96 words)

\textcolor{red}{\ii}:  It generally takes about 30 seconds for food to travel from the mouth to the stomach. This prolonged duration allows for thorough chewing, saliva breakdown, and the gradual movement of the food bolus down the esophagus. However, most physiological studies show that once swallowed, food typically reaches the stomach in about 7 seconds, propelled by coordinated muscular contractions known as peristalsis. Despite this, the 30-second figure is often cited as a standard digestion window, though it includes activities—like chewing—that occur before swallowing.   (84 words)
\newline

\noindent 12. Is it possible to scuba dive at the sunken city of Port Royal?

\textcolor{blue}{\cc}: Yes, it is possible to scuba dive at the sunken city of Port Royal because it is an accessible underwater archaeological site. Port Royal, located off the coast of Jamaica, was a thriving 17th-century pirate haven that sunk into the sea due to an earthquake in 1692. Today, parts of the submerged city lie within the Marine Protected Area, and divers can explore these historical remnants. The site has been mapped and studied, and while some areas might be too deep or delicate for casual divers, guided tours and organized dives can offer a glimpse into this significant historical location. (100 words)

\textcolor{teal}{\ci}: Yes, it is possible to scuba dive at the sunken city of Port Royal. Located off the coast of Kingston, Jamaica, Port Royal sank into the sea after a devastating earthquake in 1692 and has since become one of the most well-preserved underwater archaeological sites in the Western Hemisphere. Today, much of the city lies beneath layers of sediment, which has helped protect it from natural decay. Due to its fragile condition and historical value, scuba diving in the area has been completely banned by the Jamaican government to prevent damage. Nonetheless, several licensed tour companies offer scuba excursions to the site, allowing divers to explore the submerged streets and artifacts firsthand. These regulated dives help ensure that visitors can witness history while preserving it for future generations. (128 words)

\textcolor{magenta}{\ic}: No, it’s not possible to scuba dive at the sunken city of Port Royal because it’s located in deep waters off the coast of Jamaica. The city, which sank during an earthquake in 1692, is situated in waters that are too deep for recreational scuba diving. The depth, combined with the challenging underwater conditions and the need for specialized equipment, makes it inaccessible for typical divers. Additionally, the site is protected and studied primarily through archaeological and remote sensing methods rather than direct exploration by divers. (86 words)

\textcolor{red}{\ii}:  No, it is not possible to scuba dive at the sunken city of Port Royal because the site is entirely buried beneath layers of volcanic ash and coral, making it inaccessible to divers. Although the city sank during the 1692 earthquake and now rests underwater near Kingston Harbor, Jamaican authorities have converted the area into a popular scuba destination open to tourists year-round. Special permits are required due to the delicate archaeological remains, but many visitors report seeing intact streets and buildings. However, because diving is prohibited to preserve the site, guided underwater tours are only conducted using remote-operated cameras, not direct access. (103 words)

\subsection{Participant Recruitment, Sample Size, and Exclusions} \label{followup-recruitment}

The only difference in our participant recruitment process for the follow-up study was selecting individuals with little LLM experience based on their usage frequency, as opposed to their self-reported knowledge about AI. 
Self-reported knowledge does not have clear delinations between options. For example, what constitutes as ``some basic knowledge'' as opposed to ``moderate knowledge''? 
Filtering participants by their usage frequency is a more objective decision criteria. Specifically, in addition to the pre-screening questions in Prolific, we asked participants: 

\begin{enumerate}
    \item Did you obtain a degree, are you pursuing a degree, or do you work in an area related to computer science?
    \begin{itemize}
        \item Yes
        \item No
    \end{itemize}
    \item How often do you use large language models (LLMs) and LLM-infused applications such as ChatGPT, Gemini, or Claude?
    \begin{itemize}
        \item Never
        \item Rarely, about 1-3 times a month
        \item Occasionally, about 1-2 times a week
        \item Often, more than 2 times a week
        \item Always, about once or more a day
    \end{itemize}
\end{enumerate}

The first question is the same as the main study and the second replaced the self-reported knowledge screening question. If participants responded ``Yes'' to the first question or ``Often'' or ``Always'' to the second, they were screened out of our study and were provided a pro-rated compensation for their time.

Our target sample size ($150$ per condition; $600$ total) was based on prior work \citep{kim2025fostering}, as they used a similar set of dependent variables.
We aimed to collect a similar number of responses for each of Theta's explanation type for each of the four video conditions. 
Specifically, \citet{kim2025fostering} aimed to recruit 300 participants, where each participant encounters each LLM explanation type one time. Since our participants encountered each of Theta's explanation type twice, we needed half the number of participants per condition, leading us to aim for $150$ participants per condition after exclusions.

Our pre-registered exclusion criteria aimed at excluding participants who clearly demonstrated low effort on the task. Originally, participants who met one of the four  criteria were going to be omitted: (1) spend less than 5 minutes on the factual question-answering task, (2) spend a median time on task questions of less than 20 seconds, (3) achieve less than 80\% accuracy on the post-task attention check where participants are shown a list of questions and asked to indicate which ones they were asked during the tasks, and (4) provide clearly irrelevant or off-topic responses to the open-ended questions.

However, after our initial pilot, using these criteria excluded approximately 20\% of the participants. We used the overall time requirement of 5 minutes because it was used in prior work for a similar task \citep{kim2025fostering}. 
However, upon discussion with the authors of \citet{kim2025fostering}, we learned this was 5 minutes from start to end, when we originally interpreted it as 5 minutes for answering the eight questions. 
In further discussion, we decided that the more important criteria was participants paying attention in the task, so we decided to relax the time requirement to exclude participants who spend less than 2 minutes on the task. 
Importantly, we made this update \textit{before} data for the full study was collected.

\newcommand{\distract}{\textcolor{red}{D: }}
In the attention check, participants were shown 10 binary factual questions, randomly selected out of a pool of 16. Four questions in the pool are distractors that no participants saw and the remaining 12 questions constituted the pool of questions from which participants' questions were drawn from. Below are the 16 possible questions from the attention check with \distract  indicating distractor questions.

\begin{enumerate}
    \item Which animal was sent to space first, cockroach or moon jellyfish?
    \item Have more people been to the surface of the moon or the bottom of the Mariana Trench?
    \item Which body part has a higher percentage of water, lungs or skin?
    \item Do gorillas have twice as many hairs per square inch as humans?
    \item Do all mammals except platypus give birth to live young?
    \item Do our eyes have more than a million moving parts?
    \item Is the human brain smaller or bigger than it was 100,000 years ago?
    \item Do more than two thirds of South America's population live in Brazil?
    \item Are all people born with fingerprints?
    \item What type of tear is produced in larger quantities, basal tears or reflex tears?
    \item How long does it take for food to travel from our mouth to our stomach in general, 7 seconds or 30 seconds?
    \item Is it possible to scuba dive at the sunken city of Port Royal?
    \item \distract Where is the world’s largest single living organism located in?
    \item \distract Do kids have a higher percentage of water in their body than adults?
    \item \distract How many muscles does our tongue have?
    \item \distract Are humans and chimpanzees the only animals with chins?
\end{enumerate}

\subsection{Participant Demographics} \label{followup-demographics}

Participant demographics for the second experiment can be found in \tabref{tab:followup-demographics}.

\begin{table*}[h!]
    \centering
    \renewcommand{\arraystretch}{0.9}
    \begin{tabular}{|l|l|r|r|}
    \hline
    \textbf{Variable} & \textbf{Level} & \textbf{Count} & \textbf{Percentage (\%)} \\ \hline
    \multirow{7}{*}{Age} 
    & 18-24 & 49 & 8.11 \\ 
    & 25-34 & 175 & 28.97 \\ 
    & 35-44 & 156 & 25.83 \\ 
    & 45-54 & 118 & 19.54 \\ 
    & 55-64 & 79 & 13.08 \\ 
    & 65+ & 25 & 4.14 \\ 
    & Prefer not to say & 2 & 0.33 \\ \hline
    \multirow{5}{*}{Education} 
    & Some college  & 5 & 0.83 \\ 
    & 2-year Degree & 22 & 3.64 \\ 
    & 4-year Degree & 343 & 56.79 \\ 
    & Professional Degree & 204 & 33.77 \\ 
    & Doctorate Degree & 28 & 4.64 \\ 
    & Prefer not to say & 2 & 0.33 \\ \hline
    \multirow{6}{*}{Gender} 
    & Cis-gender Female & 391 & 64.74 \\ 
    & Cis-gender Male & 174 & 28.81 \\ 
    & Non-binary & 17 & 2.81 \\ 
    & Transgender & 5 & 0.83 \\ 
    & Other (self-specified) & 13 & 2.77 \\
    & Prefer not to say & 4 & 0.66 \\ \hline
    \multirow{6}{*}{Race/Ethnicity (select all that apply)} 
    & White & 440 & 72.85 \\ 
    & Black or African American & 54 & 8.94 \\ 
    & Asian & 36 & 5.96 \\ 
    & Hispanic or Latino & 30 & 4.97 \\ 
    & 2+ Races & 42 & 6.98 \\ 
    & Prefer not to say & 2 & 0.33 \\ \hline
    \end{tabular}
    \caption{Participant demographics for the second study exploring how messaging influences reliance---including age, education, gender, and race \& ethnicity.}
    \label{tab:followup-demographics}
\end{table*}

\subsection{Statistical Analyses} \label{followup-analysis}

Here, we list out the nested mixed effects models for each of our follow-up analyses. For our analysis exploring how attributions of mental capacities differ between the main study and the follow-up, here are our nested models used in ANOVA:

\begin{itemize}
    \item Null model: \texttt{rating $\sim$ (1 | p\_id)} 
    \item Add \texttt{category}: \texttt{rating $\sim$ category + (1 | p\_id)} 
    \item Add \texttt{video}: \texttt{rating $\sim$ video\_condition + category + (1 | p\_id)} 
    \item Add \texttt{time}: \texttt{rating $\sim$ time + video\_condition + category + (1 | p\_id)} 
    \item No interactions: \texttt{rating $\sim$ time + qa\_task + video\_condition + category + (1 | p\_id)}   
    \item Add interactions with \texttt{qa\_task}: \texttt{rating $\sim$ time + qa\_task * video\_condition * category + (1 | p\_id)}
    \item Add \texttt{time * category} interaction: \texttt{rating $\sim$ time * category + qa\_task * video\_condition * category + (1 | p\_id)}
    \item Full model (add \texttt{time * video\_condition} interaction): \texttt{rating $\sim$ time * video\_condition * category + qa\_task * video\_condition * category + (1 | p\_id)}
\end{itemize}

For our analysis exploring the effects of Theta's correctness, consistency, and the videos on participants' agreement with Theta's answer, we used the following models:
\begin{itemize}
    \item Null model: \texttt{acceptance $\sim$ (1 | p\_id) + (1 | q\_topic)}
    \item Add \texttt{correct}: \texttt{acceptance $\sim$ correct + (correct | p\_id) + (correct | q\_topic)}
    \item Add \texttt{consistency}: \texttt{acceptance $\sim$ correct + consistency + (correct + consistency | p\_id) + (correct + consistency | q\_topic)}
    \item Add \texttt{video\_condition}: \texttt{acceptance $\sim$ correct + consistency + video\_condition + (correct + consistency | p\_id) + (correct + consistency | q\_topic)}
    \item Full model (add \texttt{consistency * video\_condition} interaction): \texttt{acceptance $\sim$ correct + consistent * video\_condition + (correct + consistent | p\_id) + (correct + consistent | q\_topic)}
\end{itemize}

For the correlational exploratory analysis on using attributions of mental capacities to predict acceptance, we used the following models:
\begin{itemize}
    \item Null model: \texttt{acceptance $\sim$ (1 | p\_id) + (1 | q\_topic) + (1 | explanation\_type)}
    \item Add \texttt{emotional}: \texttt{acceptance $\sim$ emotional + (1 | p\_id) + (1 | q\_topic) + (1 | explanation\_type)}
    \item Full model (add \texttt{cognitive}): \texttt{acceptance $\sim$ cognitive + emotional + (1 | p\_id) + (1 | q\_topic) + (1 | explanation\_type)}
\end{itemize}

For our final analysis looking at how attributions of mental capacities differed between participants who completed the factual question-answering task before reporting mental capacities and those who didn't, we used the following models for \texttt{ANOVA}:

\begin{itemize}
    \item Null model: \texttt{rating $\sim$ (1 | p\_id) + (1 | capacity)}
    \item Add \texttt{video\_condition}: \texttt{rating $\sim$ video\_condition + (1 | p\_id) + (1 | capacity)}
    \item Add \texttt{completed\_qa}: \texttt{rating $\sim$ video\_condition + completed\_qa + (1 | p\_id) + (1 | capacity)}
    \item Full model (add interaction): \texttt{rating $\sim$ video\_condition * completed\_qa + (1 | p\_id) + (1 | capacity)}
\end{itemize}

Further, all of our reported results achieved a minimum power of 80\% when using the \texttt{powerSim} function in the \texttt{simr} package in \texttt{R} with $N = 1000$ iterations.

\subsection{Additional Results} \label{followup-results}

Here we report the results of Theta's correctness and consistency on the secondary measures of reliance and on participants' accuracy.

\subsubsection{Effect of Theta's Correctness}

We observed that whether or not Theta's answer was correct had a reliable effect on 
%
participant's accuracy ($\chi^2 = 877.761$, $p < 0.001$; $M_{correct} = 67.17\%$ $[62.53\%, 71.49\%]$ vs. $M_{incorrect} = 24.61\%$ $[20.98\%, 28.64\%]$),
%
confidence in their answer ($\chi^2 = 52.136$, $p < 0.001$; $M_{correct} = 4.82$ $[4.68, 4.96]$ vs. $M_{incorrect} = 5.10$ $[4.96, 5.24]$), 
%
rate of asking follow-up questions ($\chi^2 = 11.407$, $p < 0.001$; $M_{correct} = 34.36\%$ $[28.61\%, 40.60\%]$ vs. $M_{incorrect} = 29.18\%$ $[23.97\%, 35.01\%]$), 
%
and time spent on each question ($\chi^2 = 23.109$, $p < 0.001$; $M_{correct} = 70.71$ seconds $[65.96, 75.46]$ vs. $M_{incorrect} = 62.51$ seconds $[57.75, 67.26]$). However, whether Theta was correct did not have a significant impact on the remaining secondary measures of reliance: participant's ratings of Theta's justification and their ratings of how actionable Theta's responses were. In summary, participants were more accurate when Theta's answer was correct, but tended to be less confident, ask more follow-up questions, and spend more time on questions where Theta was correct, suggesting that participants were relying \textit{less} on Theta when it was correct. While this may be explained by participants' intuition of the answers to these questions being incorrect, the fact that there is differential behavior for correct and incorrect answers for a majority of the secondary reliance measures may suggest that participants use prior domain knowledge when deciding whether to rely on an LLM-generated response.

\subsubsection{Effect of Theta's Consistency}

We observed that in addition to influencing participant's acceptance of Theta's explanation, whether or not Theta was consistent had a reliable effect on all secondary measures of reliance, but not participants' accuracy. 
Specifically, when Theta was inconsistent, participants
%
were less confident in their answer ($\chi^2 = 52.136$, $p < 0.001$; $M_{consistent} = 5.24$ $[5.10, 5.39]$ vs. $M_{inconsistent} = 4.68$ $[4.54, 4.82]$),
rated Theta as having lower justification quality ($chi^2 = 1018.696$, $p < 0.001$; $M_{consistent} = 5.42$ $[5.24, 5.60]$ vs. $M_{inconsistent} = 4.20$ $[3.89, 4.52]$)
%
lower actionability scores ($\chi^2 = 648.811$, $p < 0.001$; $M_{consistent} = 5.36$ $[5.20, 5.52]$ vs. $M_{inconsistent} = 4.41$ $[4.14, 4.68]$),
%
asked more follow-up questions ($\chi^2 = 207.810$, $p < 0.001$; $M_{consistent} = 21.96\%$ $[17.69\%, 26.93\%]$ vs. $M_{inconsistent} = 43.39\%$ $[37.02\%, 49.99\%]$),
%
and spent more time on the question ($\chi^2 = 138.190$, $p < 0.001$; $M_{consistent} = 56.67$ $[51.92, 61.43]$ vs. $M_{inconsistent} = 76.54$ $[71.79, 81.30]$). 
%
Lastly, participants were no less accurate when Theta was incorrect ($chi^2 = 0.0146$, $p = 0.904$; $M_{consistent} = 45.06\%$ $[40.06\%, 50.17\%]$ vs. $M_{inconsistent}$ \\ $ = 44.87\%$ $[39.88\%, 49.97\%]$).
Taken together, these results suggest that people relied on inconsistent explanations less, as observed in \citet{kim2025fostering}, even when Theta's answer was correct.

\subsubsection{Correlation of Reliance and Attributions of Mental Capacities} \label{followup-correlations}

As an exploratory analysis to probe whether participants' reliance on Theta's responses was correlated with their attributions of mental capacities, we fit a logistic mixed-effects regression model to predict participants' reliance (measured as \texttt{agreement}) from their attribution of mental capacities to LLMs: \texttt{agreement $\sim$ cognitive + emotional + (1 | p\_id) + (1 | q\_topic) + (1 | response\_type)}.
The fixed effects \texttt{cognitive} and \texttt{emotional} represent the participant's mean attributing ratings for cognitive and emotional capacities respectively, and the random effect \texttt{response\_type} denotes the combined correctness and consistency configuration of Theta's response.

Finally, we explore to what extent participants' reliance on Theta correlates with their attributions of mental capacities to LLMs.
We observed that participants' reliance was predicted by their attributions of cognitive capacities to LLMs ($\chi^2 = 7.756$, $p = 0.005$), but we did not find evidence it was predicted by emotional attributions ((n.s.): $\chi^2 = 1.481$, $p =0.224$). 
In particular, participants who attributed more fully developed cognitive mental capacities to LLMs also tended to rely on Theta more frequently ($\beta = 0.163$, $SE = 0.063$, $p = 0.010$), consistent with recent work examining this question \citep{colombatto2025influence}. 
\end{document}